\documentclass[letterpaper,journal,final]{IEEEtran} 

\usepackage{graphicx}   
\usepackage{amssymb}
\usepackage{amsmath}
\usepackage{amsfonts}                                
\usepackage[latin1] {inputenc}
\usepackage{enumerate}
\usepackage{mathrsfs}
\usepackage{tabulary}
\usepackage{cite}
\usepackage{psfrag}
\usepackage{graphicx}          
\usepackage{color}      
\usepackage{epsfig} % for postscript graphics files
\usepackage{epstopdf}
\usepackage{times} % assumes new font selection scheme installed
\def\qedp{\hspace*{\fill}~{\tiny $\blacksquare$}}
\def\qed{\relax\ifmmode\hskip2em \Box\else\unskip\nobreak\hskip1em $\Box$\fi}

\newtheorem{theorem}{Theorem}
\newtheorem{itlemma}{Lemma}
\newtheorem{itdefinition}{Definition}
\newtheorem{itproposition}{Proposition}
\newtheorem{itresult}{Result}
\newtheorem{itremark}{Remark}
\newtheorem{itassumption}{Assumption}
\newtheorem{itcorollary}{Corollary}
\newtheorem{itexample}{Example}

\newenvironment{proposition}{\begin{itproposition}\rm}{\end{itproposition}}
\newenvironment{remark}{\begin{itremark}\rm}{\end{itremark}}
\newenvironment{assumption}{\begin{itassumption}\rm}{\end{itassumption}}
\newenvironment{lemma}{\begin{itlemma}\rm}{\end{itlemma}}

\title{\LARGE \bf Data Rates for Stabilizing Control under Denial-of-Service Attacks} 

\author{ Shuai Feng, Ahmet Cetinkaya, Hideaki Ishii, Pietro Tesi and Claudio De Persis
	\thanks{S. Feng, P. Tesi and C. De Persis are with ENTEG, Faculty of Science and Engineering, University of Groningen, 9747AG Groningen, The Netherlands ({\tt\small s.feng@rug.nl, p.tesi@rug.nl, c.de.claudio@rug.nl}).}	
	\thanks{P. Tesi is also affiliated to the DINFO, Universit\'a di Firenze, 50139 Firenze, Italy ({\tt\small pietro.tesi@unifi.it}).}	
	\thanks{ A. Cetinkaya and H. Ishii are with the Department of Computer Science, Tokyo Institute of Technology, Yokohama 226-8502, Japan ({\tt\small ahmet@dsl.mei.titech.ac.jp,  ishii@c.titech.ac.jp}).}
}

\begin{document}

\maketitle

\begin{abstract}
We study communication-constrained networked control problems for linear time-invariant systems in the presence of Denial-of-Service (DoS) attacks, namely attacks that prevent transmissions over the communication network. Our work aims at exploring the relationship between system resilience and network bandwidth capacity. Given a class of DoS attacks, we first characterize time-invariant bit-rate bounds that are dependent on the unstable eigenvalues of the dynamic matrix of the plant and the parameters of DoS attacks, beyond which exponential stability of the closed-loop system can be guaranteed. 
%On the other hand, the robustness of a networked control system under DoS and quantization is quantified, \emph{i.e.} stability can be achieved if the DoS-dependent parameter is smaller than the bit-dependent bound. 
Second, we design the time-varying bit-rate protocol and show that it can enable the system to maintain the comparable robustness as the one under the time-invariant bit-rate protocol and meanwhile promote the possibility of transmitting fewer bits especially when the attack levels are low. Our characterization clearly shows the trade-off between the communication bandwidth and resilience against DoS. An example is given to illustrate the proposed solution approach. 
\end{abstract}

\section{Introduction}
Cyber-physical systems (CPSs) have attracted much attention due to the advances in automation. Integrating communication and computation technologies, CPSs have a broad spectrum of applications ranging from small local control systems to large-scale systems, some of which are safety-critical. This raises the issue of reliability of CPSs to a considerable important level. Among a variety of aspects in reliability problems, the security of CPSs becomes a challenge from both practical and theoretical points of view. Here the concept of CPSs security mostly concerns the resilience against or protection from malicious attacks, \emph{e.g.} deceptive attacks and Denial-of-Service (DoS)\cite{cheng2017guest, teixeira2015secure}. 

This paper deals with resilient control under DoS attacks. We consider a basic problem where the network has limited bandwidth and is subject to DoS attacks, and the intention of the attacker is to induce instability. This implies that the signals transmitted over such a network are subject to both quantization and dropout. It is well known that an insufficient bit rate in the communication channel influences the stability of a networked control system\cite{antsaklis2004guest}, not to mention packet drops \cite{hespanha2007survey}. Hence, the topic of networked control under data rate constraints and random packet dropouts has been investigated by many researchers. However, those results may not be applicable in the context of DoS since the communication failures induced by DoS can exhibit a temporal profile quite different from the one induced by genuine packet losses; particularly packet dropouts induced by DoS need not follow a given class of probability distributions\cite{sastry}. This poses new challenges in theoretical analysis and controller design.  

The literature on networked control with bit-rate limitation is large and diverse\cite{brockett2000quantized, hespanha2002towards, 1310461, 867021, 1186780, 8360077, tallapragada2016event} and the problem when quantization and genuine packet losses coexist has been well studied, see \cite{5555976, okano2014stabilization, okano2017stabilization, tsumura2009tradeoffs, you2010minimum, 8052508, minero2013stabilization}.  
In \cite{1310461}, the authors obtain necessary and sufficient conditions concerning the observability and stabilization for the networked control of a linear time-invariant system under communication constraints. These conditions are independent of information patterns and only reliant on the inherent property of the considered plant, \emph{i.e.} the unstable eigenvalues of the dynamic matrix of the plant. The papers \cite{5555976, minero2013stabilization} investigate the minimum data rate problem for
mean square stability under Markovian packet losses. Necessary and sufficient conditions for stabilization are obtained for both scalar and vector systems.

Recently, systems under DoS attacks have been studied from the control-theoretic viewpoint \cite{de2015input, cetinkaya2017networked, 8049303,  ding2017multi, li2017sinr,wakaiki2017quantized, automatica2017,7526102,  senejohnny2017jamming, de2016networked, 8353464, 8036273, FENG201740, 8264578}. 
In \cite{de2015input}, a framework is introduced where DoS attacks are characterized by \emph{frequency} and \emph{duration}. 
The contribution is an explicit characterization of DoS frequency and duration 	under which stability can be preserved
through state-feedback control. Extensions have been considered dealing with self-triggered networks \cite{senejohnny2017jamming} and nonlinear systems \cite{de2016networked}. 
In \cite{cetinkaya2017networked}, the authors generalize this model and consider a scenario where malicious attacks and genuine packet losses coexist, in which the effect of malicious attacks and random packet losses are merged and characterized by an overall packet drop ratio.  
In \cite{8049303}, the authors investigate launching DoS attacks optimally to a network with genuine packet losses. Specifically, the attacker aims at maximizing the estimation error with constrained energy.  
In \cite{ding2017multi}, the authors formulate a two-player zero-sum stochastic game framework to consider a remote secure estimation problem, where the signals are transmitted over a multi-channel network under DoS attacks. 
A game-theory-based model where transmitters and jammers have multiple choices of sending and interfering power is considered in \cite{li2017sinr}.  
The recent paper \cite{wakaiki2017quantized} investigates the stabilization problem of a discrete-time output feedback system under quantization and DoS attacks. In the event of the satisfaction of a certain norm condition, a lower bound of quantization level and an upper bound of DoS duration are obtained together guaranteeing stability.

In this paper, we consider the stabilization problem of a linear time-invariant continuous process, possibly open-loop unstable and with complex eigenvalues, where the communication between sensor and controller takes place over a bit-rate limited and unreliable digital channel. Previously, we have shown that a controller with prediction capability significantly promotes the resilience of a networked control system against DoS in the sense that the missing signals induced by DoS attacks can be reconstructed and then applied for computing the control input \cite{automatica2017,7526102, FENG201740}. Under proper design, the system can achieve ISS-like robust stability or asymptotic stability in the presence or absence of disturbance and noise, respectively. However, when the network has limited bandwidth, the existing results are not applicable any longer because signal deviation induced by quantization cannot be simply treated as bounded noise, and such signal deviation influences the accuracy of estimation/prediction and hence the resilience of the closed-loop system.

Therefore, there is a trade-off between communication bandwidth and system resilience. An interesting question is to find how large the bit rate must be to ensure the stability of a system under DoS, possibly an open-loop unstable system. We may state this question in another way as how much the limited bit rate degrades the robustness of a networked control system in the context of stabilization.  
We follow the approach aligned with that for the minimum data rate control problems discussed above. In particular, we recover those results in the case without any DoS. 
Exploiting the techniques of transformation, we associate the bit rates with the eigenvalues of the dynamic matrix of the process and DoS parameters, and explicitly characterize the relationship between system resilience and bit rates. Specifically, we compute a bit-rate bound element-wise, larger than which the closed-loop system is exponentially stable. This on the other hand reveals the ``robustness degradation" induced by quantization. 
In addition, assuming that the communication protocol is acknowledgment-based, we propose a time-varying bit-rate design where the packet size is time-varying. This enables us to preserve a comparable level of robustness against DoS and meanwhile promotes the possibility of saving transmitted bits especially when the attack levels are low.

This paper is organized as follows. In Section \uppercase\expandafter{\romannumeral 2}, we introduce the framework consisting of system transformations, a class of DoS attacks and the contribution of this paper. Section \uppercase\expandafter{\romannumeral 3} is the core part of this paper. Considering that the data rate is time-invariant, we accordingly choose the uniform quantizer and design the predictor. The dynamics of quantization range and prediction error are analyzed. Then we conduct the stability analysis. In Section \uppercase\expandafter{\romannumeral 4}, the time-varying bit-rate protocol is introduced, which can stabilize the system and save communication resources. 
A numerical example is introduced in Section \uppercase\expandafter{\romannumeral 5}, and finally Section \uppercase\expandafter{\romannumeral 6} ends the paper with conclusions and possible future research directions.

\textbf{Notation}. We denote by  $\mathbb R$ the set of reals. Given $b \in \mathbb R$, $\mathbb R_{\geq b}$ and $\mathbb R_{>b}$ denote the sets of reals no smaller than $b$ and reals greater than $b$, respectively; $\mathbb R_{\le b}$ and $\mathbb R_{<b}$ represent the sets of reals no larger than $b$ and reals smaller than $b$, respectively; $\mathbb Z$ denotes the set of integers. For any $c \in \mathbb Z$, we denote $\mathbb Z_c := \{c,c+ 1,\cdots\}$. Let $\lfloor x \rfloor$ be the floor function such that $\lfloor x \rfloor= \max\{k\in \mathbb{Z}|k\le x\}$. Also, let $\lceil x \rceil$ be the ceiling function such that $\lceil x \rceil= \min\{k\in \mathbb{Z}|k \ge x\}$.
%$\lceil x \rceil$ denotes a ceiling function being $\lceil x \rceil= \min\{k\in \mathbb{Z}|k \ge x\}$.
Given a vector $\beta$, $\|\beta\|$ is its Euclidean norm.
Given a matrix $\Gamma$, $\|\Gamma\|$ represents its spectral norm and $\Gamma ^ {\text{T}}$ is its transpose. Given an interval $\mathcal{I}$, $|\mathcal{I}|$ denotes its length. The Kronecker product is denoted by $\otimes$.
Finally, given a signal $\mathcal F$, $\mathcal F(t^-)$ denotes the limit from below at $t$.

\section{Framework}

\subsection{System description}

\begin{figure}[h]
	\begin{center}
		\includegraphics[width=0.5 \textwidth]{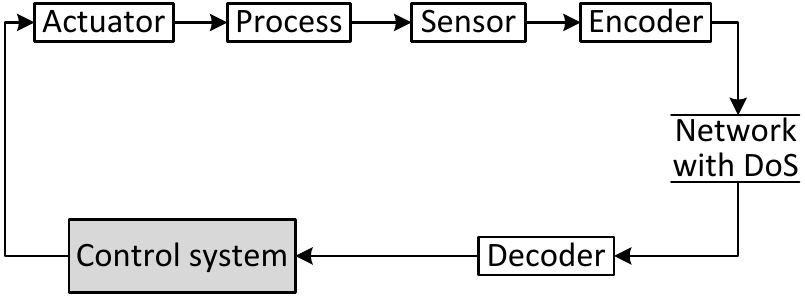} \\
		\linespread{1}\caption{Controller and actuator co-location architecture
		} \label{Figure 1}
	\end{center}
\end{figure}

Consider the networked control architecture in Figure \ref{Figure 1}. The process is a linear time-invariant continuous system given by
\begin{align}\label{vector system}
\dot x(t) = A x(t) + B u(t)
\end{align}
where $t\in \mathbb{R}_{\ge 0}$, $x(t)\in \mathbb{R}^{n_x}$ is the state with $x(0)$ arbitrary, $A\in \mathbb{R}^{n_x \times n_x}$, $B\in \mathbb{R}^{n_x \times n_u}$, $u(t) \in \mathbb{R}^{n_u}$ is the control input and $(A, B)$ is stabilizable. 
Let $K\in \mathbb{R}^{n_u \times n_x}$ be a matrix such that the real part of each eigenvalue of $A+BK$ is strictly negative.
Let $\lambda_r = c_r \pm d_r i$ be the eigenvalues of $A$ with $c_r, d_r  \in \mathbb{R}$, where $c_1, c_2, c_3,\cdots$ are distinct and $i$ represents the imaginary number. 
If $d_r = 0$ then $\lambda_r$ has only real part and corresponds to a real eigenvalue such that $\lambda_r = c_r$. If $d_r \ne 0$, $\lambda_r$ represents a pair of complex eigenvalues whose real part is $c_r$ and imaginary parts are $ d_r i$ and $- d_r i$, respectively.
In the following sections, the real part of $\lambda_r$ is denoted by $c_r$, where we do not distinguish if $\lambda_r$ is real or complex.
We assume that the state is measurable by sensors. 

The measurement channel has limited bandwidth and is moreover subject to DoS attacks. The transmission attempts between the encoder and decoder are carried out periodically with interval $\Delta$, \emph{i.e.} 
\begin{align}\label{periodic transmission interval}
t_{k+1}-t_k=\Delta
\end{align}
where $\{t_k\}_{k\in \mathbb{Z}_0}=\{t_0, t_1, \cdots\}$ denotes the sequence of the instants of transmission attempts. By convention, we let $t_0=0$. Moreover, we assume that the network communication protocol is acknowledgment-based (like the TCP protocol) without any delay in terms of both encoded signal and acknowledgment transmissions.

Since we consider a controller-actuator co-location architecture (\emph{cf.} Figure \ref{Figure 1}), only the measurement channel is subject to DoS, and the control channel is free from DoS disruptions and always available. Due to DoS attacks, not all the transmission attempts succeed. Hence, we denote by $\{z_m\}_{m\in \mathbb{Z}_0}=\{z_0, z_1, \cdots\}   \subseteq \{t_k\}_{k\in \mathbb{Z}_0} $ the sequence of the time instants at which successful transmissions occur.

\subsection{System transformation}
In order to facilitate the analysis in Sections \uppercase\expandafter{\romannumeral 3} and  \uppercase\expandafter{\romannumeral 4}, we carry out two transformations in this subsection.

First, we transform the original process (\ref{vector system}) into the real Jordan canonical form.
Let $S\in \mathbb{R}^{n_x \times n_x}$ be a transformation matrix such that (\ref{vector system}) can be rewritten as 
\begin{align}\label{Jordan system}
\dot{\tilde{x}}(t) = \tilde{A} \tilde{x} (t) + \tilde{B} u (t)
\end{align}
where $\tilde{x}(t) = S x(t)$, $t\in \mathbb{R}_{\ge 0}$ and $\tilde{A}\in \mathbb{R}^{n_x \times n_x} $ is the Jordan form of $A$ such that 
\begin{align}\label{Jordan form}
\tilde{A}= S A S^{-1}=  \text{diag}(A_1, A_2, \cdots, A_p), \,\,p\in \mathbb{Z}_1
\end{align}
%\begin{align}\label{Jordan form}\setlength{\arraycolsep}{2pt}
%\bar{A}= S A S^{-1}=  \left[\begin{array}{cccc}
%A_1 & & & \\ & A_2 &   & \\ & & \ddots & \\ & & & A_p 
%\end{array}\right], \,\,p\in \mathbb{Z}_1.
%\end{align}
in which $p$ represents the number of Jordan blocks. Let $r = 1, 2, \cdots, p$.
The Jordan block associated with the real eigenvalue $\lambda_r=c_r$ is
\begin{align}\label{Jordan block}\setlength{\arraycolsep}{3pt}
A_r = \left[\begin{array}{cccc}
c_r &1& & \\ & c_r & 1 & \\ & & \ddots &1 \\ & & & c_r 
\end{array}\right] \in  \mathbb{R}^{n_r \times n_r}
\end{align}
where $n_r$ is the order of $A_r$. 
The Jordan block associated with the complex eigenvalues $\lambda_r= c_r \pm d_r i$ ($d_r \ne 0$) is
\begin{align}\label{Jordan complex block}\setlength{\arraycolsep}{3pt}
A_r = \left[\begin{array}{cccc}
D_r &I& & \\ & D_r & I & \\ & & \ddots &I \\ & & & D_r 
\end{array}\right] \in  \mathbb{R}^{2n_r \times 2n_r}
\end{align}
with 
\begin{align}\setlength{\arraycolsep}{3pt} \label{Dr and I}
D_r=
\left[\begin{array}{cc}
c_r & -d_r \\ d_r & c_r  
\end{array}\right], \,\,
I=
\left[\begin{array}{cc}
1 & 0\\ 0 & 1
\end{array}\right]
\end{align}
where $2n_r$ is the order of $A_r$\cite{perko2013differential}.
Meanwhile we have $\tilde{B}=SB\in \mathbb{R}^{n_x \times n_u} $. 
If $A$ has only real eigenvalues, the Jordan form of $A$ in (\ref{Jordan form}) with the Jordan blocks in (\ref{Jordan block}) is sufficient for further development. However, in the event of the existence of complex eigenvalues in $A$, 
we need one more step of transformation, which is carried out by the lemma below.

\begin{lemma}\label{Second transformation}
	Consider the process in (\ref{Jordan system}) where $\tilde A$ is in the Jordan form as in (\ref{Jordan form}). There exists a transformation $\bar x(t) = E(t) \tilde{x}(t)$ such that (\ref{Jordan system}) can be transformed into 
	\begin{align} \label{second Jordan system}
	\dot {\bar x} (t) = \bar A  \bar x(t) + \bar B(t) u(t) 
	\end{align} 
	where
	\begin{align}
	\bar A &= E(t) \tilde{A} E(t)^{-1} + \dot E(t) E(t)^{-1} \nonumber\\
	&= \left[\begin{array}{cccc}
	\bar A_1 & & & \\ & \bar A_2 &   & \\ & & \ddots & \\ & & & \bar A_p 
	\end{array}\right], \,\,p\in \mathbb{Z}_1
	\end{align}
	with 
	\begin{align}\setlength{\arraycolsep}{3pt} \label{second Jordan block}
	\bar A_r = A_r = \left[\begin{array}{cccc}
	c_r &1& & \\ & c_r & 1 & \\ & & \ddots &1 \\ & & & c_r
	\end{array}\right] \in  \mathbb{R}^{n_r \times n_r}
	\end{align}
	corresponding to the real eigenvalue $\lambda_r =c_r$, and 
	\begin{align}\setlength{\arraycolsep}{3pt} \label{second Jordan block complex}
\bar A_r  = \left[\begin{array}{cccc}
c_r &1& & \\ & c_r & 1 & \\ & & \ddots &1 \\ & & & c_r
\end{array}\right] \otimes I \in  \mathbb{R}^{2n_r \times 2n_r}
\end{align}
%	\begin{align}\setlength{\arraycolsep}{3pt} \label{second Jordan block complex}
%	\bar A_r = 
%	\left[\begin{array}{cccc}
%	c_r I  & I & & \\ & c_r I  & I & \\ & & \ddots & I \\ & & & c_r I 
%	\end{array}\right] \in \mathbb{R}^{2 n_i \times 2n_i }
%	\end{align}
	corresponding to the complex eigenvalues $\lambda_r=c_r \pm d_r i$ with $d_r \ne 0$. Besides, $\bar B(t) = E(t) \tilde{B}$.
\end{lemma}

\emph{Proof.} We refer the readers to the Appendix for the proof including the design of $E(t)$. \qedp

In \cite{1310461} and \cite{mazenc2011interval}, similar techniques of transformation where the transformation matrix is time-varying are used. 
It is trivial to mention that one can directly transform (\ref{vector system}) into (\ref{second Jordan system}) by computing $\bar A = E(t) S A S^{-1} E(t)^{-1} + \dot E(t) E(t)^{-1}$ and $\bar B(t) = E(t) S B$. In case of the existence of complex eigenvalues, $E(t)$ is a time-varying matrix. This implies that $\bar B(t)$ is time-varying.

\subsection{Time-constrained DoS}
We refer to DoS as the phenomenon for which some transmission attempts may fail.
We consider a general DoS model
that constrains the attacker action in time 
by only posing limitations on the frequency of DoS attacks and their duration. Let 
$\{h_n\}_{n \in \mathbb Z_0}$ with $h_0 \geq 0$ denote the sequence 
of DoS \emph{off/on} transitions, that is,
the time instants at which DoS exhibits 
a transition from zero (transmissions are successful) to one 
(transmissions are not successful).
Hence,
\begin{eqnarray}  \label{DoS_intervals}
H_n :=\{h_n\} \cup [h_n,h_n+\tau_n[  
\end{eqnarray}
represents the $n$-th DoS time-interval, of a length $\tau_n \in \mathbb R_{\geq 0}$,
over which the network is in DoS status. If $\tau_n=0$, then
$H_n$ takes the form of a single pulse at $h_n$.  
Given $\tau,t \in \mathbb R_{\geq0}$ with $t\geq\tau$, 
let $n(\tau,t)$
denote the number of DoS \emph{off/on} transitions
over $[\tau,t]$, and let 
\begin{eqnarray}  \label{DoS_intervals_union}
\Xi(\tau,t) := \bigcup_{n \in \mathbb Z_0} H_n  \, \bigcap  \, [\tau,t] 
\end{eqnarray}
be the subset of $[\tau,t]$ where the network is in DoS status. 

\begin{assumption}
	(\emph{DoS frequency}). 
	There exist constants 
	$\eta \in \mathbb R_{\geq 0}$ and 
	$\tau_D \in \mathbb R_{> 0}$ such that
	\begin{align} \label{ass:DoS_slow_frequency} 
	n(\tau,t)  \, \leq \,  \eta + \frac{t-\tau}{\tau_D}
	\end{align}
	for all  $\tau,t \in \mathbb R_{\geq0}$ with $t\geq\tau$.
	\qedp
\end{assumption}

\begin{assumption} 
	(\emph{DoS duration}). 
	There exist constants $\kappa \in \mathbb R_{\geq 0}$ and $T  \in \mathbb R_{>1}$ such that
	\begin{align} \label{ass:DoS_slow_duration}
	|\Xi(\tau,t)|  \, \leq \,  \kappa + \frac{t-\tau}{T}
	\end{align}
	for all  $\tau,t \in \mathbb R_{\geq0}$ with $t\geq\tau$. 
	\qedp
\end{assumption}

\begin{remark}
	Assumptions 1 and 2
	do only constrain a given DoS signal in terms of its \emph{average} frequency and duration.
	Following \cite{hespanha1999stability}, 
	$\tau_D$ can be defined as the average dwell-time between 
	consecutive DoS off/on transitions, while $\eta$ is the chattering bound.
	Assumption 2 expresses a similar 
	requirement with respect to the duration of DoS. 
	It expresses the property that, on the average,
	the total duration over which communication is 
	interrupted does not exceed a certain \emph{fraction} of time,
	as specified by $1/T$.
	Like $\eta$, the constant $\kappa$ plays the role
	of a regularization term. It is needed because
	during a DoS interval, one has $|\Xi(h_n,h_n+\tau_n)| = \tau_n >  \tau_n /T$.
	Thus $\kappa$ serves to make (\ref{ass:DoS_slow_duration}) consistent. 
	Conditions $\tau_D>0$ and $T>1$ imply that DoS cannot occur at an infinitely
	fast rate or be always active. \qedp
\end{remark}

The next lemma relates DoS parameters and the time elapsing between successful transmissions.
\begin{lemma}\label{Lemma Q}
	Consider periodic transmission attempts $t_{k+1}-t_k=\Delta$,
	along with a DoS attack satisfying Assumptions 1 and 2. If 
	$\frac{1}{T}+\frac{\Delta}{\tau_D}<1$, then the sequence of successful transmissions
	satisfies $z_0 \leq Q$ and 
	$z_{m+1}-z_{m} \leq Q + \Delta$ for all $m \in \mathbb Z_0$, 
	where
	\begin{align}
	Q := (\kappa + \eta \Delta) \left(1-\frac{1}{T} - \frac{\Delta}{\tau _D} \right)^{-1}.
	\end{align}
\end{lemma}

\emph{Proof}.
The proof of Lemma 1 in \cite{FENG201740} carries over to this lemma with $s_0$, $s_{r+1}$ and $s_r$ replaced by $z_0$, $z_{m+1}$ and $z_m$, respectively.  
\qedp \\

The following lemma presents the relationship between DoS parameters, time and the number of successful transmissions therein. 

\begin{lemma} \label{Lemma T}
Consider the DoS attacks characterized by Assumptions 1 and 2. The number of successful transmissions within the interval $[z_0, z_m[$, which is denoted by $T_S(z_0, z_m) $, satisfies 
	\begin{align}
	T_S(z_0, z_m)
	\ge \frac{1-\frac{1}{T}-\frac{\Delta}{\tau_D}}{\Delta} (z_m - z_0) - \frac{\kappa+\eta\Delta}{\Delta} 	   	
	\end{align} 
	where $z_m \ge z_0$ and $\Delta$ is as in (\ref{periodic transmission interval}).	
\end{lemma}

\emph{Proof.}
Consider an interval $[z_0, z_m]$ with $z_m\ge z_0 $ and let $H_n$ represent the $n$-th DoS time-interval within $[z_0, z_m]$ here. One can verify that the number of unsuccessful transmissions during $H_n$ is no larger than $ \frac{\tau_n}{\Delta}+1$.
%The duration of $\tilde \Xi (\tau, t)$ satisfies $|\tilde \Xi (\tau, t)| \le |\Xi (\tau, t)| + n(\tau, t)\Delta$.
%\begin{align}
%|\tilde \Xi (\tau, t)| \le |\Xi (\tau, t)| + n(\tau, t)\Delta
%\end{align} 
Hence the number of unsuccessful transmissions during $[z_0, z_m]$ satisfies
\begin{align}\label{Tu}
T_U(z_0, z_m) & \le \sum_{k=0}^{n(z_0, z_m)-1}( \frac{\tau_k}{\Delta}+1) \nonumber\\
&\le  \frac{|\Xi (z_0, z_m)|}{\Delta}  +  n(z_0, z_m)
\end{align} 
Let $T_A(z_0, z_m) =\frac{z_m-z_0}{\Delta} +1$ denote the number of total transmission attempts during $[z_0, z_m]$. Note that $T_A(z_0, z_m)$ and $T_S(z_0, z_m)$ are defined corresponding to the intervals $[z_0, z_m]$ and $[z_0, z_m[$, respectively.
Therefore it follows from (\ref{ass:DoS_slow_frequency}), (\ref{ass:DoS_slow_duration}), and (\ref{Tu}) that $T_S(z_0, z_m) $ satisfies
\begin{align}\label{Ts}
T_S(z_0, z_m) &= T_A(z_0, z_m) - T_U(z_0, z_m) -1 \nonumber\\
&\ge \frac{1-\frac{1}{T}-\frac{\Delta}{\tau_D}}{\Delta} (z_m-z_0 ) - \frac{\kappa+\eta\Delta}{\Delta} 	   	
\end{align}
%where $T_S(z_0, z_m)$ represents the number of successful transmissions during $[z_0, z_m[$.
\qedp

\begin{remark}
In the scenario of a reliable network ($T=\tau_d=\infty$ and $\kappa=\eta=0$), $Q$ in Lemma \ref{Lemma Q} becomes zero, and $T_U(z_0, z_m)=0$ implies $T_S(z_0, z_m) =  T_A(z_0, z_m)-1$. This means that every transmission attempt ends up with a successful transmission. Thus, Lemmas \ref{Lemma Q} and \ref{Lemma T} describe the functioning of a standard periodic transmission policy. 
\qedp
\end{remark}

\subsection{Literature review}
The robustness problem of the structure as in Figure \ref{Figure 1} has been investigated in \cite{automatica2017} and \cite{7526102}, where we assumed the network has infinite bandwidth and the measurements are not quantized. For the ease of comparison and clarifying the contribution of this paper, we briefly recall the controller and the result in \cite{7526102}. 
The control system is given by
{\setlength\arraycolsep{2pt} 
\begin{eqnarray}  \label{co-located predictor}
	\arraycolsep=1.4pt\def\arraystretch{1.7}
	\left\{ \begin{array}{l}
	u(t) = K \xi (t) \\
	\left\{ \begin{array}{ll}
	\dot{\xi }(t)  = A  {\xi}(t)+B u(t) , & \quad \textrm{if } t\ne z_m \\
	\xi(t)=x(t)+n(t), & \quad \textrm{if } t=z_m
	\end{array} \right. 
	\end{array} \right.
\end{eqnarray}
where $\xi(t)$ is the estimation of $x(t)$ and $n(t)$ represents bounded noises. 
\begin{theorem}\cite{7526102}
	Consider the dynamical system as in (\ref{vector system}) under a co-located control system as in 
	(\ref{co-located predictor}). 
	The closed-loop system is stable
	for any DoS sequence 
	satisfying Assumptions 1 and 2 with
	arbitrary $\eta$ and $\kappa$, and with $\tau_D$ and $T$ such that
	\begin{eqnarray} \label{co-located controller bound of DoS}
	\frac{1}{T} + \frac{\Delta}{\tau_D} \, < \,  1
	\end{eqnarray} 
	\qedp
\end{theorem}

It is trivial that in case $n(t)=0$, the result above still holds.

\subsection{Paper contribution}
%The control objective of this paper is to explore the fundamental relationship between communication bandwidth and system's resilience against DoS attacks. From designing a robust system point of view, it is interesting to investigate how large or redundant the bit rate must be in order to stabilize a system under potential DoS attacks. On the other hand, if the bit rate of a network is given, we would like to characterize the maximum amount of DoS attacks (the value of $\frac{1}{T}+ \frac{\Delta}{\tau_D}$) under which closed-loop stability is preserved. 
Exploiting the controller in (\ref{co-located predictor}) and the architecture in Figure \ref{Figure 1}, we first design the encoder and decoder such that they are free of over-flow of quantization range even in the presence of DoS attacks. After fixing the control system's structure, the number of bits $R_r$ for coding is the only parameter to be taken care of. Given the control framework, the contribution of this paper is mostly in finding the appropriate $R_r$, possibly under the presence of DoS attacks.

The main contribution of this paper is two-fold.
\begin{itemize}
\item[i)] 

The first contribution is to show that the closed-loop system is exponentially stable if the time-invariant bit rate $R_r$ satisfies
\begin{align}\label{contribution 1}
R_r\left \{
\begin{array}{ll}
> \frac {1}{1-\frac{1}{T}- \frac{\Delta}{\tau_D}} c_r \Delta \log_2 e, &  \textrm{if} \,\, c_r \ge 0 \\
\ge 0,   &  \textrm{if}\,\, c_r < 0
\end{array}\right. 
\end{align}
where $R_r$ represents the number of bits applied to the signals corresponding the $r$-th block in $\bar A$. The condition (\ref{contribution 1}) is general enough in the sense that in the absence of DoS attacks, the result of minimum data rate control is recovered (\emph{cf.} Remark \ref{Remark 5}). 
On the other hand, we characterize the robustness of the system, namely the amount of DoS attacks less than which stability can be still preserved. One can preserve closed-loop stability if the frequency and duration of DoS attacks satisfy 
\begin{align}\label{Contribution robustness}
\frac{1}{T} + \frac{\Delta}{\tau_D} < 1- \frac{c_r \Delta \log_2 e}{R_r}, \,\,\, \forall c_r \ge 0
\end{align}
where $R_r > 0$. Clearly, the signal inaccuracy due to quantization cannot be simply treated as the one caused by measurement noises in the sense that the noises do not enter the right-hand side of (\ref{co-located controller bound of DoS}), whereas the quantization degrades the system's robustness by diminishing the right-hand side of (\ref{co-located controller bound of DoS}) into (\ref{Contribution robustness}). This implies that some DoS attacks that used to be tolerable would now cause instability. 

\emph{}
\item[ii)] 
As a second contribution, we propose the time-varying bit-rate protocol consisting of \emph{bit-computing parts} and \emph{coding parts} in both the encoding and decoding systems. The bit-computing parts are able to generate sequences of time instants. By resorting to using acknowledgments,  the sequences of time instants can be synchronized in the encoding and decoding systems. Based on the generated time sequences (under the influences of DoS), the number of bits applied for each transmission attempt can be pre-determined. If the DoS attack is short, a number of bits no larger than $R_r$ could guarantee the decay of quantization range, and there is no need to apply $R_r$, which leads to the possibility of saving bits. Under suitable choices of parameters, we show that the closed-loop system is stable if the maximum number of bits that the encoding and decoding systems can apply in one transmission attempt satisfies (\ref{contribution 1}).

\end{itemize}

\section{Time-invariant bit rate}
In this section, we introduce the design of the encoding and decoding systems, and the control system, where the number of bits used for coding are time-invariant. 

\subsection{Uniform quantizer}
The limitation of bandwidth implies that transmitted signals are subject to quantization. Let 
\begin{align}
\chi_l:= \frac{e_l}{j_l}
\end{align}
be the original $l$-th signal before quantization and
$q_{\mathcal R_l} (\chi_l)$ represents the quantized signal of $\chi_l$ with $\mathcal R_l$ bits, where $l= 1, 2, 3, \cdots, n_x$. The choices of $\mathcal R_l$, $e_l \in \mathbb{R}$ and $j_l\in\mathbb{R}_{ > 0}$ will be specified later. We implement a uniform quantizer such that 
\begin{align} \label{uniform quantizer}
q_{\mathcal R_l} (\chi_l) : = 
\left \{ \begin{array}{ll} 
\frac{\lfloor 2^{\mathcal R_l-1} \chi_l \rfloor + 0.5}{ 2^{\mathcal R_l-1} }, &  \quad  \textrm{if } -1 \le \chi_l < 1  \\ 
1- \frac{0.5}{2^{\mathcal R_l-1}}, & \quad  \textrm{if } \chi_l=1
\end{array} \right. 
\end{align} 
if $\mathcal R_l \in \mathbb{Z}_1 $ and 
\begin{align}\label{uniform quantizer 2}
q _{\mathcal R_l} (\chi_l) =0
\end{align}
if $\mathcal R_l = 0 $. Note that for any $j_l\in \mathbb{R}_{> 0} $ the following property holds:
\begin{align} \label{Property of quantizer}
\left| e_l - j_l q_{\mathcal R_l} \left(\frac{e_l }{j_l}\right)  \right| \le \frac{j_l}{2^{\mathcal R_l}}, \quad \textrm{if}   \,\frac{|e_l|}{j_l} \le 1
\end{align}
for both cases, namely $\mathcal R_l \in \mathbb{Z}_0 $ \cite{5555976, you2010minimum}. For the ease of visualizing (\ref{uniform quantizer}), Figure \ref{Figure quantizer} shows the quantization function with $\mathcal R_l=2$. The quantizer applied in the time-varying bit-rate protocol will be presented in Section \uppercase\expandafter{\romannumeral 4}.

\begin{figure}[th]
	\begin{center}
		\includegraphics[width=0.27 \textwidth]{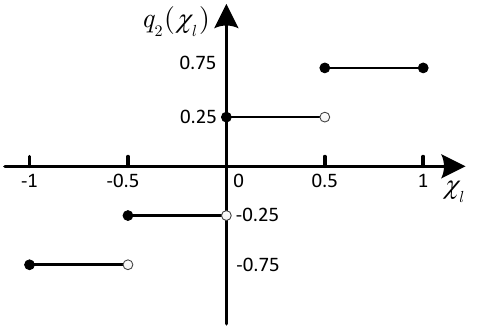} \\
		\linespread{1}\caption{Example of quantization with $\mathcal R_l = 2$. For instance, any number falling into $[0, 0.5[$ would be quantized into 0.25. 
		} \label{Figure quantizer}
	\end{center}
\end{figure}

\subsection{Control architecture}

The basic idea of the control system design is that we equip the encoding and decoding systems with prediction capability  to properly quantize data and more importantly predict the missing signals that are interrupted by DoS. Specifically, the encoding system outputs quantized signals and transmits them to the decoding system through a DoS-corrupted network. The decoding system attempts to predict future signals based on the received quantized signals.
Notice that the following design is based on $\dot{\bar{x}}(t) = \bar{A} \bar{x} (t) + \bar{B}(t) u (t)$. 

As shown in Figure \ref{Figure 2}, on the sensor side the encoding system is embedded with a predictor for predicting $\bar x(t)$. Let $\hat x(t)=[\hat x_1(t)\,\, \hat x_2(t)\,\, \cdots \,\, \hat x_{n_x}(t)]^\text{T}$ denote the prediction of $\bar{x}(t) = [\bar x_1(t)\,\, \bar x_2(t)\,\, \cdots \,\, \bar x_{n_x}(t)]^\text{T}$. The error $e(t)=[e_1(t)\,\, e_2(t)\,\, \cdots \,\, e_{n_x}(t)]^\text{T}$ describes the discrepancy between $\bar{x}(t)$ and $\hat{x}(t)$, where
\begin{eqnarray}\label{e error definition} 
e_l(t):=\hat x_l(t)- \bar{x}_l(t), \,\,\, l= 1, 2, \cdots, n_x. 
\end{eqnarray}
Furthermore, we will design a dynamic system (see (\ref{Jordan case J})-(\ref{Jordan case H}) below), whose state $J(t)= [j_1(t) \,\,\, j_2(t)\,\,\, \cdots \,\,\, j_{n_x}(t)]^\text{T}$ is always positive.
Namely, it is $j_l(t) > 0$ for $t\in \mathbb{R}_{\ge 0}$, where $j_l(t)$ represents the quantization range that bounds the error, \emph{i.e.} $j_l(t) \ge  |e_l(t)|$ for $t\in \mathbb{R}_{\ge 0}$ as it will be shown in the next subsection.  
Recalling that $\chi_l(t) : = \frac{e_l(t)}{j_l(t)}$, $j_l(t) \ge|e_l(t)|$ for $t\in \mathbb{R}_{\ge 0}$ implies $|\chi_l(t)|  = \frac{|e_l(t)|}{j_l(t)} \le 1$ for $t\in \mathbb{R}_{\ge 0}$, and hence there is no overflow problem and the quantizer (\ref{uniform quantizer})-(\ref{uniform quantizer 2}) is valid for $t\in \mathbb{R}_{\ge 0}$. More importantly, $j_l(t) \ge|e_l(t)|$ for $t\in \mathbb{R}_{\ge 0}$ would make (\ref{Property of quantizer}) hold for $t\in \mathbb{R}_{\ge 0}$. 

On the actuator side, the decoding system is a copy of the encoding system. Once there is a successful transmission containing the encoded state, it recovers $q_{\mathcal R_l}(\chi_l(z_m))$ based on the received code and updates the predictor embedded in the decoding system, and sends an acknowledgment back to the encoding system. The acknowledgment would enable the encoding system to know the successful transmission reception. We assume that the encoding and decoding systems have the same initial conditions. Therefore, identical structures and initial conditions, and acknowledgments would guarantee synchronization of all the signals in the encoding and decoding systems.

The predictor in both the encoding and decoding systems predicting $\bar x(t)$ is given by
\begin{align}\label{Jordan predictor}
\left\{ \begin{array}{ll} 
\dot{\hat x}(t) = \bar{A} \hat x(t) + \bar B(t) u(t), & t\ne z_m \\
\hat x(t) = \hat x(t^-) - \Phi(t^-), & t=z_m.
\end{array}\right.
\end{align}
As for the input $u(t)$, we have $u(t)= \bar K(t) \hat x(t)$
%\begin{align}\label{Jordan controller}
%u(t)= \bar{K} \hat x(t),
%\end{align}
where $\bar K(t)= K S^{-1} E(t)^{-1} \in \mathbb{R}^{n_u \times n_x }$. In the encoding system, $u(t)$ is applied only to the predictor. In the decoding system, $u(t)$ is applied to both the predictor and actuator (see Figure \ref{Figure 2}).

The column vector $\Phi(t)$ in (\ref{Jordan predictor}) is given by
\begin{align}\label{Jordan case Phi}
\Phi(t)=\left [  \begin{array}{ccc}
\phi_1(t) \\
\vdots\\
\phi_{n_x}(t)
\end{array}  \right]=\left [  \begin{array}{ccc}
j_1 (t) q_{\mathcal  R_1}( \chi_1(t) ) \\
\vdots\\
j_{n_x} (t) q_{\mathcal  R_{n_x}}( \chi _{n_x}(t) )
\end{array}  \right]
\end{align}
where $\chi_l(t)= \frac{e_l(t)}{j_l(t)}$ and $j_l(t)$ is the $l$-th entry in the column vector $J(t)= [j_1(t) \,\,\, j_2(t)\,\,\, \cdots\,\,\, j_{n_x}(t)]^\text{T}$, which is the solution to the impulsive system
\begin{eqnarray} \label{Jordan case J}
\left \{ \begin{array}{ll}
\dot J(t) = \bar{A}  J(t), & t\ne z_m \\
J(t) =H J(t^-), & t= z_m
\end{array}\right.
\end{eqnarray} 
with
%\begin{eqnarray}\label{Jordan case H}
%H \in \mathbb{R}^{n_x \times n_x}= \left [ \begin{array}{cccc}
%2^{-R_1}I_1 & & &\\ & 2^{-R_2}I_2 &  &\\ & & \ddots &\\  & & & 2^{-R_p}I_p  
%\end{array}\right]
%\end{eqnarray}
\begin{align}\label{Jordan case H}
H = \text{diag}(2^{-R_1}I_1, 2^{-R_2}I_2, \cdots, 2^{-R_p}I_p) \in \mathbb{R}^{n_x \times n_x}
\end{align}
where $I_r \in \mathbb{R}^{n_r\times n_r}$ or $I_r \in \mathbb{R}^{2n_r\times 2n_r}$ represents an identity matrix corresponding to $\bar A_r$ in (\ref{second Jordan block}) or (\ref{second Jordan block complex}), respectively. At the moment of a successful transmission, $J(t)$ in both the encoding and decoding systems is updated according to the second equation in (\ref{Jordan case J}). At last,
the initial conditions of $\hat x $ and $J$ in the encoding and decoding systems are identical and satisfy
\begin{align} \label{Jordan initial condition}
\left \{ \begin{array} {l}
\hat x_l (0^- ) = 0,  \\
j_l(0^-) >  |\bar{x} _l(0^-)|, 
\end{array} \right.  l= {1, 2, \cdots, n_x}
\end{align} 

%The entire procedure 
%\itemize
%\item[1.] At each $t_k$, the estimator computes $\hat x(t_k)$ and then obtains the error $e(t_k)$.
%\item[2.] Take the value of $j_l(t_k)$ and obtain $\chi_l(t_k)$, and quantized it into $q_{R_l}(\chi_l(t_k)))$ and sends the codes. 
%\item[3.] 

%\begin{figure}[t]
%	\begin{center}
%		\includegraphics[width=0.5 \textwidth]{Drawing6.pdf} \\
%		\linespread{1}\caption{Control architecture. The controller is equipped with prediction capability, whose prediction is based on the quantized signal. One the process side, a copy of the controller is implemented in order to compute the prediction error $e(t)$, which is reset when an acknowledgement is received.  
%		} 
%	\end{center}
%\end{figure}

\begin{figure}[t]
	\begin{center}
		\includegraphics[width=0.49 \textwidth]{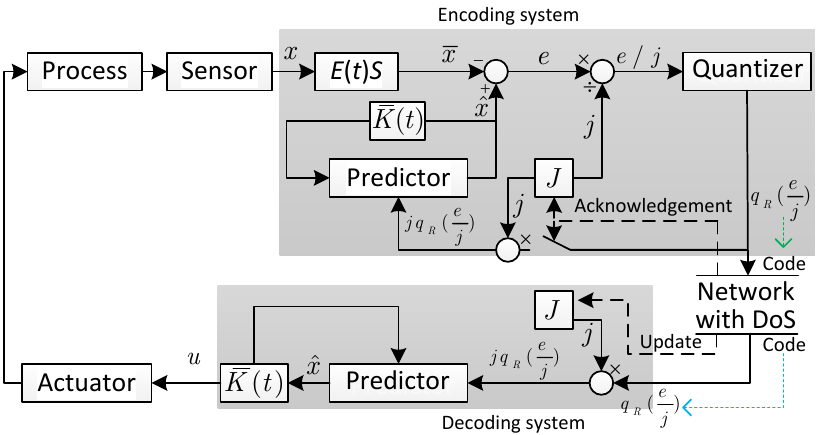} \\
		\linespread{1}\caption{Control architecture with encoding and decoding systems. The black solid lines and dashed lines represent paths of signals computed by embedded blocks and triggering signals generated by communication protocol, respectively. The green dashed line represents the process that converts signals into code and the blue one represents the reversed process.  
		} \label{Figure 2}
	\end{center}
\end{figure}

It is worth mentioning that $\mathcal R_l$ represents the number of bits applied to the $l$-th quantized signal, which is element-wise based. Since the $l$-th quantized signal must be associated with one block $\bar A_r$ ($r=1, 2, \cdots, p$), therefore, in this paper the data rate analysis is based on the index of $\bar A_r$, and all the elements corresponding to $\bar A_r$ would apply $R_r$ bits. For example, if the $l$-th signal is associated with $\bar A_r$, then $\mathcal R_l = R_r$. In the results of this paper, we will obtain the bounds of $\{R_r\}_{r=1, 2, \cdots, p}$, so that $\{\mathcal R_ l\}_{l=1, 2, \cdots, n_x}$ can be determined.

\subsection{Overflow-free quantizer }

In this part, our intention is to show that $j_l(t) \ge|e_l(t)|$ for $t\in \mathbb{R}_{\ge 0}$ with $l=1, 2, \cdots, n_x$. Exploiting (\ref{e error definition})-(\ref{Jordan case Phi}) and the continuity of $\bar{x}(t)$ such that $\bar{x}(t) = \bar{x}(t^-)  $, we have
\begin{align}\label{Jordan e_i and j_i at z_m}
e_l(t) &= \hat x _l (t) -\bar{x}_l(t)   \nonumber \\  
&= \hat  x_l (t^-) -\bar{x}_l(t^-) - j_l(t^-) q_{\mathcal R_l} \left(\frac{e_l(t^-)}{j_l(t^-)} \right),\,\, t=z_m \nonumber\\
&=  e_l(t^-) - j_l(t^-) q_{\mathcal R_l} (\frac{e_l(t^-)}{j_l(t^-)})
\end{align}
where $l= {1, 2, \cdots, n_x}$. 
Hence the dynamics of $e(t)$ obeys
\begin{align} \label{Jordan error dynamics}
\left \{ \begin{array} {ll}
\dot e(t) = \bar {A} e(t), & t \ne z_m \\
e(t) = e(t^-) - \Phi(t^-) , & t = z_m
\end{array} \right.
\end{align}

Moreover, observing $e(t)$ in (\ref{Jordan error dynamics}) and $J(t)$ in (\ref{Jordan case J}), one has
\begin{align}
\begin{array}{c}
\dot e(t) = \bar {A} e(t), \\
\dot J(t) = \bar {A} J(t),
\end{array}  \,\,\,\,\,\, t\ne z_m
\end{align}
whose solutions are $e(t) = e^{\bar A t} e(0)$ and $J(t) = e^{\bar A t} J(0)$, respectively, for $0 \le t < z_0$ (if $z_0 \ne 0$) or $0 \le t < z_1$ (if $z_0 = 0$), where
\begin{align}
e^{\bar A t} = \text{diag}(U_1(t), U_2(t), \cdots, U_p(t))
\end{align}
with
\begin{align}\label{U_r}
U_r(t) =e^{c_r t}  V_r(t) \otimes W, \,\,\,\, r=1, 2, \cdots, p
\end{align}
where 
\begin{align}\label{V_r}
V_r(t) = \left[\begin{array}{ccccc}
1&  t  &   \cdots& \cdots & \frac{t^{n_r-1}}{(n_r-1)!} \\ & 1 & t & \cdots & \frac{t ^{n_r-2}}{(n_r-2)!}  \\  && \ddots  & \ddots &\vdots \\&&  & \ddots&t \\ &&&&1 
\end{array}\right]
\end{align}	
and
\begin{align}\label{R_r}
W= \left\{
\begin{array}{lc}
1, & \text{if}\,\, d_r = 0	
\\
I, & \text{if}\,\, d_r \ne 0
\end{array}  \right.
\end{align}
in which $I$ is as in (\ref{Dr and I}).

By $e(t) = e^{\bar A t} e(0)$, one can obtain that $|e(t)| \le e^{\bar A t} |e(0)|$ holds element-wise, where $|\quad|$ denotes a function that computes the absolute value of each element in a vector, \emph{i.e.} $|e(t)|=[|e_1(t)|\,\, |e_2(t)|\,\, \cdots\,\, |e_{n_x}(t)|]^{\text T}$.
Define the column vector $\varepsilon(t) : = J(t) - |e(t)|= [\varepsilon_1(t) \,\, \varepsilon_2(t) \,\,\cdots\,\, \varepsilon_{n_x}(t)]^{\text{T}}$. 
If $z_0 \ne 0$, one has
\begin{align}\label{J and e 0}
\varepsilon(t) &=  J(t) - |e(t)|  \nonumber\\
&\ge   e^{\bar A t} J(0) -  e^{\bar A t} |e(0)|, \,\,\,\,\, 0\le t < z_0 \nonumber\\	
&= e^{\bar A t} \varepsilon(0).
\end{align}
By (\ref{Jordan initial condition}), one knows that 
\begin{align}
\varepsilon(0)    &= J (0) - |e(0)| \nonumber\\
					&= J(0)  - |\hat x(0) - \bar x(0)| \nonumber\\
					&= J(0^-)  - |\hat x(0^-) - \bar x(0^-)| \nonumber\\
					&= J(0^-)  - |\bar x(0^-)|
\end{align}
and thus every element in the column vector $\varepsilon(0)$ is positive, which implies that every element in the column vector $\varepsilon(t)$ is positive for $0\le t < z_0$. Thus, one can infer that $j_l(z_0 ^-)- |e_l(z_0 ^-)|  >  0 $, and hence $j_l(z_0 ^-) - |e_l(z_0 ^-)|  \ge  0 $. 
In view of (\ref{Jordan e_i and j_i at z_m}), it is clear that 
\begin{align}\label{no over flow e and j}
|e_l(z_0)| &=   \left| e_l(z_0 ^-) - j_l(z_0 ^ -) q_{\mathcal R_l} \left(\frac{e_l(z_0 ^-)}{j_l(z_0 ^-)}\right) \right|  \nonumber\\
&\le \frac{j_l (z_0 ^-)}{2^{\mathcal R_l}}= j_l(z_0) 
\end{align}
where the inequality is implied by (\ref{Property of quantizer}) and the second equality in (\ref{Jordan case J}), from which one obtains that $|e_l(z_0)| \le j_l(z_0)  $ and furthermore $\varepsilon(z_0) \ge 0$. Following the analysis as in (\ref{J and e 0}), one could obtain that $\varepsilon(t) \ge  e^{\bar A (t - z_0) } \varepsilon(z_0)$ with $z_0\le t < z_1$.
%	\begin{align}
%	\varepsilon(t) = e^{\bar A t} \varepsilon(z_0), \,\, z_0\le t < z_1
%	\end{align}
This implies that every element in $\varepsilon(t)$ is non-negative and $|e_l(t)| \le  j_l(t) $ for $z_0 \le t < z_1$. 
By simple induction, we can verify that	$|e_l(t)| \le j_l(t)$ for $t\in \mathbb{R}_{\ge0}$
if $z_0 \ne 0$. 

If $z_0 = 0$, we know that $|e_l(z_0^-)| = |e_l(0^-)| = |\bar x_l(0^-)| < j_l(0^-)= j_l(z_0 ^-)$, and hence $j_l(z_0 ^-) - |e_l(z_0 ^-)|  \ge  0 $. 
Following (\ref{no over flow e and j}), one gets $|e_l(z_0)|\le j_l(z_0)$. The remaining part follows the same analysis as in the  scenario $z_0  \ne 0$ to obtain $|e_l(t)| \le j_l(t)$ for $t\in \mathbb{R}_{\ge0}$ when $z_0 = 0$. Therefore, we conclude that
\begin{align}\label{Jordan e_i < j_i}
|e_l(t)| \le j_l(t), \,\,\, l=1, 2, \cdots, n_x,  \,\,\,\, t\in \mathbb{R}_{\ge0}
\end{align} 
and thus the quantizer (\ref{uniform quantizer}) does not undergo any overflow, and (\ref{Property of quantizer}) always holds.
Notice that (\ref{Jordan e_i < j_i}) holds for $t\in \mathbb{R}_{\ge0}$, which implies $|e_l(t)|$ is always bounded by $j_l(t)$ in the absence or presence of DoS attacks. 
Without losing generality, we focus the attention from $z_0$ onwards.

%One sees that the encoding and decoding systems integrating the predictors and the uniform quantizers guarantee that there is no overflow problem even under the presence of DoS attacks. The rational that we transmit quantized $\chi_l$ instead of quantized $x_l$ is due to the constraints imposed by the uniform quantizer, \emph{i.e.} $|\chi_l|\le 1$ in (\ref{uniform quantizer}). Otherwise, one would need $|x_l|\le 1$ for $t\in \mathbb R_{\ge 0}$, which would be easily violated under DoS attacks causing overflow problems.

\subsection{Dynamics of the encoding and decoding systems}
Since the evolutions of the signals in the encoding and decoding systems are identical, we would present this part from the view of either the encoding or the decoding system, and omit the other one. 

Considering the impulsive system (\ref{Jordan case J})--(\ref{Jordan case H}), we obtain that
\begin{align}\label{Jordan j_i(z_m) and j_i(z_m-1)}
J(z_m) 
&=H e^{\bar{A}(z_m-z_{m-1})} J(z_{m-1})	\nonumber\\
&=P(z_{m-1}, z_m) J(z_{m-1})
\end{align}
where 
\begin{align}
&P(z_{m-1}, z_m) \nonumber\\
= & \mbox{\,\,}  H e^{\bar{A}(z_m-z_{m-1})} \nonumber\\
=& \mbox{\,\,} \text{diag}(P_1(z_{m-1}, z_m), P_2(z_{m-1}, z_m), \cdots, P_p(z_{m-1}, z_m)).
\end{align}
%\begin{align}
%&P(z_{m-1}, z_m) \nonumber\\
% =&
%\left [ \begin{array}{cccc}
%P_1(z_{m-1}, z_m)	 & \mathbf{0} & \mathbf{0} & \mathbf{0} \\ \mathbf{0} &   P_2(z_{m-1}, z_m)	 & \mathbf{0} & \mathbf{0}  \\ \mathbf{0}&\mathbf{0} & \ddots & \mathbf{0}\\  \mathbf{0}& \mathbf{0}& \mathbf{0}& P_p(z_{m-1}, z_m)
%\end{array}\right]
%\end{align}
%where $\mathbf{0}$ is a matrix with all entry equaling zero.
Note that $P(z_{m-1}, z_m)$ is a block diagonal matrix in which 
\setlength{\arraycolsep}{2pt}
\begin{align} \label{P_r first time}
P_r(z_{m-1}, z_m) 
&= 2^{-R_r}  U_r(\Delta_m) 
\end{align}
with $r=1,2, \cdots, p$, $\Delta_m = z_m -z_{m-1}$ and 
$U_r(\Delta_m)$ can be obtained from (\ref{U_r}).

%and $\otimes$ represents the Kronecker product.
%Notice that $P_r(z_{m-1}, z_m) \in \mathbb{R}^{n_r \times n_r}$ if the eigenvalue corresponding to the Jordan block $A_r$ is real such that $\lambda_r =\lambda_r + d_r i$ with $d_r=0$. Otherwise $P_r(z_{m-1}, z_m)  \in \mathbb{R}^{2n_r \times 2n_r} $ in case $\lambda_r =\lambda_r + d_r i$ with $d_r \ne 0$. 

Iteratively from (\ref{Jordan j_i(z_m) and j_i(z_m-1)}), we obtain that
\begin{align}\label{Jordan J(zm) and J(z0):1}
J(z_m)  
&= \prod_{k=1}^{m} P(z_{k-1}, z_{k}) J(z_{0}) \nonumber \\
&= P(z_0, z_m) J(z_{0})
\end{align}
where $P(z_{0}, z_m) := \prod_{k=1}^{m} P(z_{k-1}, z_{k}) $ is a block diagonal matrix given by $P(z_{0}, z_m) 
= \text{diag}(P_1(z_{0}, z_m), P_2(z_{0}, z_m), \cdots, P_p(z_{0}, z_m))$
%\begin{align}
%&P(z_{0}, z_m) \nonumber\\
% =& \text{diag}(P_1(z_{0}, z_m), P_2(z_{0}, z_m), \cdots, P_p(z_{0}, z_m))
%\end{align}
%\setlength{\arraycolsep}{2pt}
%\begin{align}\label{Jordan T(z_0, z_m)}
%&P(z_{0}, z_m) \in \mathbb{R}^{n_x \times n_x}  \nonumber\\
%=&
%\left [ \begin{array}{cccc}
%P_1(z_{0}, z_m) &  &  & \\ 
%&   P_2(z_{0}, z_m) &  &  \\ & & \ddots & \\  & & & P_p(z_{0}, z_m)
%\end{array}\right]
%\end{align}
in which
\begin{align} \label{P_r before proof}
P_r(z_{0}, z_m)=\prod_{k=1}^{m} P_r(z_{k-1}, z_{k}) 
\end{align}
with $r=1,2,\cdots, p$. 

Recall that $\{z_m\}_{m\in \mathbb{Z}_0}$ denotes the sequence of time instants of the successful transmissions. Now we introduce a lemma concerning the convergence of $J(z_{m})$. 
\begin{lemma}\label{Jordan Lemma convergence of l} 
	Consider the dynamics of $J(t)$ in (\ref{Jordan case J})--(\ref{Jordan case H}) and the DoS attacks in Assumptions 1 and 2 satisfying $\frac{1}{T}+ \frac{\Delta}{\tau_D}<1$ with $\Delta$ being the sampling interval of the network as in (\ref{periodic transmission interval}).  All the elements in the column vector $J( z_{m})$ converge to zero as $z_m\to \infty$ if $R_r$ satisfies 
	\begin{align} \label{Jordan constant data rate for stability}
	R_r\left \{
	\begin{array}{ll}
	> \frac {1}{1-\frac{1}{T}- \frac{\Delta}{\tau_D}} c_r \Delta \log_2 e, &  \textrm{if} \,\, c_r \ge 0 \\
	\ge 0 ,  &  \textrm{if}\,\, c_r < 0
	\end{array}\right. r=1, 2, \cdots, p
	\end{align}
	where $c_r$ is the real part of $\lambda_r$.
\end{lemma}

\emph{Proof.} 
In this proof, we mainly show that $\|P(z_{0}, z_m)\|$ converges to zero as $z_m\to \infty$ if $\frac{1}{T}+ \frac{\Delta}{\tau_D}<1$ and (\ref{Jordan constant data rate for stability}) are satisfied, which implies the convergence of $J(z_m)$.

According to (\ref{P_r first time}) and (\ref{P_r before proof}), we have
\begin{align}\label{P_r}
P_r(z_{0}, z_m) &=\prod_{k=1}^{m} P_r(z_{k-1}, z_{k})  \nonumber\\
&= \prod_{k=1}^{m}  (2^{-R_r}  U_r(\Delta_k)   )                                 \nonumber\\
&=  (2^{-R_r}  )^m   U_r(\sum_{k=1}^{m}\Delta_k) .  
\end{align}
Substituting (\ref{U_r}) into (\ref{P_r}), we obtain
\begin{align} \label{convergence 1}
 &P_r(z_0, z_m) \nonumber\\
=& \,\,  \frac{e^{c_r(z_m -z_0)}}{(2^{R_r})^m}  V_r(z_m -z_0) \otimes W \nonumber\\
=&  \,\,  \frac{e^{c_r(z_m -z_0)}  (z_m -z_0)^{n_r-1} }{ (2^{R_r})^m} \frac{V_r(z_m -z_0)}{ (z_m -z_0)^{n_r-1} }  \otimes W. 									  	
\end{align}
It is easy to verify that 
\begin{align}\label{convergence 2}
&\frac{V_r(z_m -z_0)}{ (z_m -z_0)^{n_r-1}} \otimes W \nonumber\\
\le & 
\left[\begin{array}{ccccc}
\frac{1}{(z_m -z_0) ^{n_r -1}}  &   \frac{1}{(z_m -z_0) ^{n_r -2}}  &   \cdots& \cdots & 1 \\ &   \frac{1}{(z_m -z_0) ^{n_r -1}} &  \frac{1}{(z_m -z_0) ^{n_r -2}} & \cdots & \frac{1 }{(z_m -z_0)}  \\  && \ddots  & \ddots &\vdots \\&&  & \ddots& \frac{1}{(z_m -z_0) ^{n_r -2}} \\ &&&&  \frac{1}{(z_m -z_0) ^{n_r -1}}
\end{array}\right] \nonumber\\
& \otimes W \nonumber\\
\le &
\left[\begin{array}{ccccc}
\frac{1}{\Delta ^{n_r -1}}  &   \frac{1}{\Delta ^{n_r -2}}  &   \cdots& \cdots & 1 \\ &   \frac{1}{\Delta ^{n_r -1}} &  \frac{1}{\Delta ^{n_r -2}} & \cdots & \frac{1 }{\Delta}  \\  && \ddots  & \ddots &\vdots \\&&  & \ddots& \frac{1}{\Delta ^{n_r -2}} \\ &&&&  \frac{1}{\Delta ^{n_r -1}}
\end{array}\right] \otimes W
\end{align}
which is upper bounded for $z_m -z_0 \ge \Delta$. 
Meanwhile, exploiting that $m= T_S(z_0, z_m)$ in Lemma \ref{Lemma T}, we have
\begin{align}\label{Jordan J(zm) and J(z0):2}
&\frac{e^{c_r(z_m -z_0)}  (z_m -z_0)^{n_r-1} }{ (2^{ R_{r}})^m} \nonumber\\
=& \,\, \frac{e^{c_r(z_m - z_0)}}{(2^{R_r})^{T_S(z_0, z_m)}}    (z_m -z_0)^{n_r-1}  \nonumber\\
\le& \,\, \theta_r  \left(\frac{e^{c_r}}{2^{ R_r {\frac{1-\frac{1}{T}- \frac{\Delta}{\tau_D}}{\Delta}}}}\right)^{z_m-z_0}   (z_m -z_0)^{n_r-1}
\end{align}
where $\theta_r:=2^{\frac{R_r(\kappa+\eta\Delta)}{\Delta}}    $.
If (\ref{Jordan constant data rate for stability}) holds and $\frac{1}{T}+\frac{\Delta}{\tau_D}<1$, it is simple to verify that
\begin{align} \label{alpha i}
\alpha _r: = \frac{e^{ c_r}}{2^{ R_r{\frac{1-\frac{1}{T}- \frac{\Delta}{\tau_D}}{\Delta}}}}  <1 .
\end{align}
This implies that there exist a finite number $C_0 ^r$ and $\mu_r <0$ such that 
\begin{align}\label{convergence 3}
&\frac{e^{c_r(z_m -z_0)}  (z_m -z_0)^{n_r-1} }{ (2^{ R_{r}})^m} \nonumber\\
\le&\,\, \theta_r  (\alpha_r)^{z_m-z_0}   (z_m -z_0)^{n_r-1} \nonumber\\
\le&\,\,  C_0 ^r e^{\mu_r (z_m -z_0)}.
\end{align}

In view of (\ref{convergence 1}), (\ref{convergence 2}) and (\ref{convergence 3}), there exists a finite $C_1^r$ such that
\begin{align}
\|P_r(z_0 ,z_m)\| &\le  C_0^r e^{\mu_r (z_m -z_0)} \left\| \frac{V_r(z_m -z_0)}{ (z_m -z_0)^{n_r-1} }  \otimes W  \right\| \nonumber\\
				  &\le  C_1 ^r e^{\mu_r (z_m -z_0)}
\end{align}
and hence we obtain that there exists finite $C_2$ and $\mu$ such that
\begin{align} \label{convergen J(zm)}
\|J(z_m)\|  \le C_2 e^{\mu (z_m -z_0)}  \| J(z_0) \| .
\end{align}
Finally we obtain the convergence of $J(z_m)$ when $z_m \to \infty$.
\qedp

%\begin{remark}
%It is worth mentioning that the states corresponding to $\lambda_r = 0$ are special in the sense that they need at least one bit to be encoded and guarantee the convergence of $j_r(z_m)$. \qedp
%\end{remark}

After proving the convergence of $J(z_m)$, now we introduce another lemma concerning the convergence of $J(t)$ and $e(t)$.

\begin{lemma}\label{Jordan lemma J(t) and e(t)}
	Consider $J(t)$ and $e(t)$ whose dynamics are given by (\ref{Jordan case J})-(\ref{Jordan case H}) and (\ref{Jordan error dynamics}), respectively. Suppose that the DoS attacks in Assumptions 1 and 2 satisfy $\frac{1}{T}+ \frac{\Delta}{\tau_D}<1$. If the bit rate $R_r$ satisfies (\ref{Jordan constant data rate for stability}), then $J(t)$ and $e(t)$ converge exponentially to the origin. 
\end{lemma}

\emph{Proof.}
According to (\ref{Jordan case J}), (\ref{convergen J(zm)}) and Lemma \ref{Lemma Q}, for $z_m \le t < z_{m+1}$, we have 
\begin{align} \label{Jordan J(t) and J(z_0)}
\|J(t)\| 
& \le e^{\bar v(t- z_m)} \|J(z_m)\| \nonumber\\
& \le e^{v(z_{m+1} - z_m)} \|J(z_m)\| \nonumber\\
& \le C_2 e^{v(z_{m+1} - z_m)}    e^{\mu (z_m -z_0)}  \| J(z_0) \|   \nonumber\\
&  =  C_2 e^{v(z_{m+1} - z_m)}    e^{\mu (z_{m+1} -z_0  + z_m - z_{m+1}  )}  \| J(z_0) \|   \nonumber\\
&  =  C_2 e^{v(z_{m+1} - z_m)}    e^{-\mu (z_{m+1} -z_m)}    e^{\mu (z_{m+1} -z_0)}  \| J(z_0) \| \nonumber\\
&  \le  C_2 e^{v(Q+\Delta)}    e^{-\mu (Q+\Delta)}    e^{\mu (z_{m+1} -z_0)}  \| J(z_0) \| \nonumber\\
&  \le  \gamma_0 e^{\mu(t-z_0)}  \| J(z_0) \|
\end{align} 
where $v=\max\{0, \bar v\}$ with $\bar v= \lambda_{\max} (\frac{\bar A + \bar A^{\text{T}}}{2})$ denoting the logarithmic norm of $\bar A$ and $\gamma_0 :=  C_2 e^{v(Q+\Delta)} e^{-\mu (Q+\Delta)}$. Since $ \gamma_0$ is finite and $\mu < 0$, we conclude that $J(t)$ exponentially converges to the origin when $t \to \infty$. 
In light of (\ref{Jordan e_i < j_i}), one could also obtain  
\begin{align}\label{Jordan e(t) and J(z_0)}
\|e(t)\| \le \|J(t)\| \le  \gamma_0 e^{\mu (t -z_0)} \|J(z_0)\| 
\end{align}
which implies the convergence of $e(t)$.
\qedp

\subsection{Main result}

Now we are ready to present the main result of this paper. 

\begin{theorem} \label{Jordan stability condition via constant rate}
	Consider the linear time-invariant process (\ref{vector system}) and its transformed system (\ref{second Jordan system}) with control action  (\ref{Jordan predictor})-(\ref{Jordan initial condition}) under the transmission policy in (\ref{periodic transmission interval}). The transmitted signals are quantized by the uniform quantizer (\ref{uniform quantizer})-(\ref{uniform quantizer 2}). Suppose that the DoS attacks characterized in Assumptions 1 and 2 satisfy $\frac{1}{T}+ \frac{\Delta}{\tau_D}<1$. If the bit rate $R_r$ with $r=1, 2, \cdots, p$ satisfies (\ref{Jordan constant data rate for stability}) then the state of the closed-loop system exponentially converges to the origin.
\end{theorem}

\emph{Proof.} 
Recall the control input $u(t)=\bar K(t) \hat x(t) = K S^{-1} E(t)^{-1} \hat x(t)= K \hat x_p(t)$, where $\hat x_p(t) = S^{-1} E(t)^{-1} \hat x(t)$ can be interpreted as the estimation of the original process state $x(t)$ in (\ref{vector system}). Then one has the error between the estimation of $x(t)$ (\emph{i.e.} $\hat x_p(t)$) and $x(t)$ such that $e_p(t):= \hat x_p(t) - x(t)$. Thus (\ref{vector system}) can be rewritten as 
$\dot x(t) = (A+BK)x(t)+ BK e_p(t)$,
whose solution is
\begin{align} \label{solution}
x(t) =&\,\, e^{(A+BK)(t-z_0)} x(z_0) + \int_{z_0}^{t}  e^{(A+BK)(t-\tau) }  BK  e_p(\tau)  d\tau  
\end{align}
where $t\in \mathbb{R}_{\ge z_0}$. 
From the equation above, one sees that the stability of $x(t)$ depends on $e_p(t)$. Thus, we analyze $e_p(t)$ such that 
\begin{align}
e_p(t)  &= \hat x_p(t) - x(t) \nonumber\\
		&=  S^{-1} E(t)^{-1} \hat x(t) - S^{-1} E(t) ^ {-1} \bar x(t) \nonumber\\
		&=  S^{-1} E(t)^{-1} (\hat x(t) - \bar x(t)) \nonumber\\
		&=  S^{-1} E(t)^{-1} e(t).
\end{align}

If $\frac{1}{T}+ \frac{\Delta}{\tau_D}<1$ and $R_r$ satisfies (\ref{Jordan constant data rate for stability}), then (\ref{Jordan e(t) and J(z_0)}) holds. Then one has
\begin{align}\label{original error}
\|e_p(t) \| &\le  \|S^{-1} E(t)^{-1} \| \|e(t)\| \nonumber\\
			&\le  \|S^{-1} E(t)^{-1} \|  \,\, \gamma_0 e^{\mu (t-z_0)} \|J({z_0})\| \nonumber\\
			&\le  \gamma_1 e^{\mu (t-z_0)} \|J({z_0})\|.  
\end{align}
Note that such $\gamma_1$ exists and is finite since $\|S^{-1} E(t)^{-1} \|$ is bounded. 
Taking the norm of both sides of the solution (\ref{solution}) and applying (\ref{original error}), one has
\begin{align} \label{exponential stability of state}
\|x (t)\| 
\le& \,\,  e^{\sigma(t-z_0)}\| x (z_0)\| +  \int_{z_0}^{t}  e^{\sigma (t-\tau) }  \|BK\|  \| e_p(\tau)\|  d\tau \nonumber\\ 
\le& \,\,  e^{\sigma(t-z_0)}\| x (z_0)\|  \nonumber\\  
& +  \int_{z_0}^{t}  e^{\sigma (t-\tau) }\|BK\| \gamma_1 e^{\mu (\tau -z_0)} \|J({z_0})\|    d\tau \nonumber\\  
\le& \,\,  e^{\bar \xi (t-z_0)}\| x (z_0)\|  \nonumber\\  
& +  \int_{z_0}^{t}  e^{\bar \xi (t-\tau) }\|BK\| \gamma_1 e^{\bar \xi (\tau -z_0)} \|J({z_0})\|    d\tau \nonumber\\  
\le& \,\,  e^{\bar \xi (t-z_0)}\| x (z_0)\|  \nonumber\\  
& +  (t-z_0) e^{\bar \xi (t -z_0)} \gamma_1 \|BK\|   \|J({z_0})\|   
\end{align}
where $\sigma<0$ is the logarithmic norm of $A+BK$ and $\bar \xi :=  \max \{\mu , \sigma\} \in \mathbb R_{<0}$. Since $\bar \xi <0$, there exist two finite reals $\delta $ satisfying $\bar \xi < \delta < 0$ and $C_3$ such that $(t-z_0) e^{\bar \xi (t -z_0)} \le C_3 e^{\delta (t-z_0)}$. Then we have
\begin{align}
\|x (t)\|  \le e^{\delta (t-z_0)} (\|x(z_0)\| + C_3 \gamma _ 1 \|BK\| \|J(z_0)\|).
\end{align}
It is immediate to see that $x(t)$ exponentially converges to the origin as $t\to \infty$. 

Moreover, in view of (\ref{Jordan J(t) and J(z_0)}) and (\ref{Jordan e(t) and J(z_0)}), and the fact that $\bar x(t)= E(t) S x(t)$ and $\|\hat x(t)\|\le \|e(t)\| + \|\bar x(t)\|$, we conclude that $J(t)$, $e(t)$, $\bar x(t)$, $\hat x(t)$ and $x(t)$ exponentially converge to the origin as $t\to \infty$. This completes the proof.   
\qedp

\begin{remark} \label{Remark 4}
We emphasize that this theorem characterizes how the bit rate influences the system's resilience. Condition (\ref{Jordan constant data rate for stability}) can be rewritten as 
	\begin{align}\label{Jordan DoS condition for constant data rate}
	\frac{1}{T} + \frac{\Delta}{\tau_D} < 1- \frac{
		c_r \Delta \log_2 e}{R_r}, \,\,\, \forall c_r\ge 0
	\end{align}
	where $R_r > 0$. The inequality above explicitly quantifies how the data rate affects the robustness, \emph{e.g.} the larger $R_r$, the smaller $T$ and $\tau_D$ can be, which implies that the system can tolerate more DoS attacks in terms of duration and frequency, and still preserve stability. Figure \ref{zone of stability} exemplifies this characterization. \qedp
	\begin{figure}[h]
		\begin{center}
			\includegraphics[width=0.4 \textwidth]{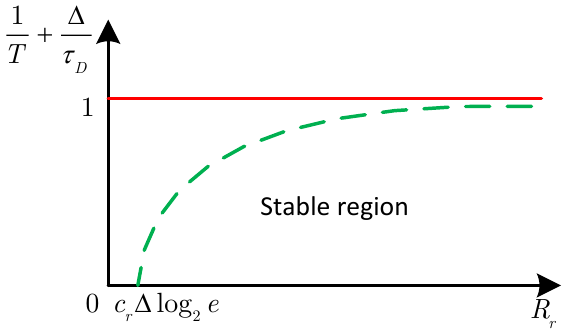} \\
			\linespread{1}\caption{Characterization of system resilience and data rate. The green dashed curve is the function $	\frac{1}{T} + \frac{\Delta}{\tau_D} = 1- \frac{
					c_r \Delta \log_2 e}{R_r}$ with $c_r > 0$. If the pair $(R_r, \frac{1}{T}+\frac{\Delta}{\tau_D})$ is in the stable region (strictly under the green dashed curve), then the system is stable. If $c_r=0$, the stable region is in rectangular shape. 
			}\label{zone of stability} 
		\end{center}
	\end{figure}
\end{remark}
\begin{remark}\label{Remark 5}
	In view of Theorem \ref{Jordan stability condition via constant rate}, if the network is reliable ($T=\tau_D = \infty$ and $\kappa = \eta = 0$), one obtains that the closed-system is exponentially stable if $R_r$ satisfies 
	\begin{align} \label{no DoS data rate}
	R_r\left \{
	\begin{array}{ll}
	> c_r \Delta \log_2 e, &  \textrm{if} \,\, c_r \ge 0 \\
	\ge 0 ,  &  \textrm{if}\,\, c_r < 0
	\end{array}\right. r=1, 2, \cdots, p.
	\end{align}
	To this end, we almost recover the result (Theorem 6) obtained in \cite{hespanha2002towards}, where no attacks were considered. By ``almost", we mean that if one omits the disturbance and noise, and considers asymptotic stabilization in \cite{hespanha2002towards}, then the data rate in (\ref{no DoS data rate}) and Theorem 6 in \cite{hespanha2002towards} are equivalent and minimum, namely they are necessary and sufficient conditions. This is the advantage of the result achieved in this paper in the aspect of recovering the minimum data rate, comparing with the one considering output-feedback scenario in \cite{wakaiki2017quantized}.  \qedp
\end{remark}

Under Theorem \ref{Jordan stability condition via constant rate}, the average data rate associated with the successfully received packets is
\begin{align}\label{average data rate encoder}
D_d:=&\lim\limits_{z_{m}  \to \infty}\frac{\sum_{l=1}^{n_x} \mathcal R_l T_S(z_0, z_{m})   }{z_{m} - z_0} \nonumber\\
>& \mbox{\,\,} \sum_{k=\{l|c_l\ge 0\}}^{} c_k \log_2 e
\end{align} 
which essentially depends on the real parts  of the eigenvalues of the dynamic matrix of the process. The average data rate associated with the transmission attempts is
\begin{align}\label{average data rate decoder}
D_e:=&\lim\limits_{z_{m} \to \infty}\frac{ \frac{z_{m}-z_0}{\Delta}  \sum_{l=1}^{n_x} \mathcal R_l  }{z_{m} - z_0}\nonumber\\
=&\,\,  \frac{\sum_{l=1}^{n_x} \mathcal R_l}{\Delta} \nonumber\\
>&\,\,   \frac{1}{1 - \frac{1}{T} - \frac{\Delta}{\tau_D}}  \sum_{k=\{l|c_l\ge 0\}}^{} c_k \log_2 e
\end{align} 
which is the corresponding result on packet size under DoS attacks comparing with the achieved result in \cite{tatikonda2004control} where genuine packet dropout is considered. 
Moreover, under a 100\% reliable network, one should have $D_e=D_d$, in which case what is sent by the encoder is fully received by the decoder. Due to the presence of DoS attacks, a larger average bit rate associated with transmission attempts is needed, namely $D_e > D_d$, and the lower bound of the average data rate associated with transmission attempts is scaled by $\frac{1}{1-\frac{1}{T}-\frac{\Delta}{\tau_D}} \in \mathbb{R}_{\ge 1}$ in (\ref{average data rate decoder}). This reflects the need of redundant communication resources to compensate for the side effect of DoS attacks.

\subsection{Stability condition over the average data rate}
We have shown that if Theorem \ref{Jordan stability condition via constant rate} holds, then the closed-loop system is stable. The setting there is that the number of bits transmitted at $z_m$ ($m=0, 1, \cdots$) are identical and equivalent to $R_r$. 
In this subsection, we loosen the sufficient condition above in the sense that the number of bits transmitted at each successful transmission time ($z_m$) does not have to be identical. Later we will show that if the average value of them is greater than $\frac{1}{1-\frac{1}{T}-\frac{\Delta}{\tau_D}}c_r \Delta \log_2e$ with $c_r \ge 0$, then the closed-loop system is still stable.

Assume that the number of bits assigned to each transmission attempt is arbitrary, and let $R_r(t_k)$ denote the number of bits applied to each element corresponding to $\bar A_r$ at $t_k$. Notice that $R_r(t_0), R_r(t_1), \cdots$ are not necessarily identical.  
Due to the physical constraints of communication equipments, it is practical to assume that the maximum number of bits that the network can transmit in one transmission is finite, namely $R_r(t_k) < \infty$ for $k \in \mathbb{Z}_0$. 
This implies that the average value $\tilde{R}_{r,k}:=\frac{R_r(t_0) + R_r (t_1) + \cdots + R_r (t_{k-1})}{k} < \infty$. It is easy to verify that $\{R_r(z_m)\} \subseteq \{R_r(t_k)\}$ and $\bar{R}_{r,m}:=\frac{R_r(z_0) + R_r (z_1) + \cdots + R_r (z_{m-1})}{m} < \infty$ for $m=1, 2, \cdots$.

Recall the definition of $\{t_k\}_{k \in \mathbb{Z}_0}$ and $\{z_m\}_{m \in \mathbb{Z}_0}$. The proposition below presents the sufficient condition for stability concerning the average data rate. 

\begin{proposition} \label{stability average data rate}
	Under the transmission policy in (\ref{periodic transmission interval}), consider the process (\ref{vector system}) and its transformed system (\ref{second Jordan system}) with control action (\ref{Jordan predictor})-(\ref{Jordan initial condition}) and the uniform quantizer (\ref{uniform quantizer})-(\ref{uniform quantizer 2}), where $\mathcal R_l=R_r(t_k)$ are arbitrary and finite at each $t_k$. The DoS attacks are characterized as in Assumptions 1 and 2 and satisfy $\frac{1}{T}+ \frac{\Delta}{\tau_D}<1$.
	 If the average value of bits along $\{z_{m-1}\}_{m=1, 2, \cdots}$ satisfies
	\begin{align}\label{average stability}
    \bar	R_{r,m}\left \{
\begin{array}{ll}
>\frac{1}{1-\frac{1}{T}-\frac{\Delta}{\tau_D}} c_r \Delta \log_2 e, &  \textrm{if} \,\, c_r \ge 0 \\
\ge 0 ,  &  \textrm{if}\,\, c_r < 0
\end{array}\right. r=1, 2, \cdots, p
	\end{align}
then the closed-loop system is stable.	 
\end{proposition}

\emph{Proof.} By observing (\ref{P_r}), we could obtain that $P_r(z_0, z_m)$ under the average data rate scenario is given by
$P_r(z_0, z_m) = U_r (\sum_{k=1}^{m} \Delta_k) \prod_{k=1}^{m} 2^{-{R_r(z_{k-1})}} $.
Then we have
\begin{align} \label{P_r average}
& P_r(z_0, z_m) \nonumber\\
=&\,\,  U_r (\sum_{k=1}^{m} \Delta_k) \prod_{k=1}^{m}  2^{-{R_r(z_{k-1})}}  \nonumber\\
=& \,\,  \frac{e^{c_r(z_m -z_0)}}{\prod_{k=1}^{m} 2^{{R_r(z_{k-1})}}}  V_r(z_m -z_0) \otimes W \nonumber\\
=& \,\,  \frac{e^{c_r(z_m -z_0)}  (z_m -z_0)^{n_r-1}              }{ (2^{\bar R_{r,m}})^m} \frac{V_r(z_m -z_0)}{ (z_m -z_0)^{n_r-1}       }  \otimes W 									  	
\end{align}
Exploiting that $m= T_S(z_0, z_m)$ in Lemma \ref{Lemma T}, we have
\begin{align}\label{sub P_r average}
&  \frac{e^{c_r(z_m -z_0)}  (z_m -z_0)^{n_r-1}              }{ (2^{\bar R_{r,m}})^m} \nonumber\\
\le &  \,\, \bar \theta_{r,m} (\bar \alpha_{r, m})^{z_m - z_0} (z_m - z_0)^{n_r -1} 
\end{align}
where $\bar \theta_{r,m} : = 2^{\frac{\bar R_{r,m} (\kappa + \eta \Delta) }{\Delta}}$ is finite and
\begin{align}
\bar \alpha_{r,m} := \frac{c_r}{2^{\bar R_{r,m} \frac{1-\frac{1}{T}- \frac{\Delta}{\tau_D}}{\Delta}}}< 1
\end{align}
if (\ref{average stability}) holds. The rest of the proof can follow the analysis after (\ref{alpha i}), and we obtain the stability of the closed-loop system. 
\qedp

%In view of the proofs of Lemma \ref{Jordan lemma J(t) and e(t)} and Theorem \ref{Jordan stability condition via constant rate}, the convergence of $J(t)$, $e(t)$ $\bar x(t)$, $\hat x(t)$ and $x(t)$ can be obtained. \qedp

It is worth mentioning that Proposition \ref{stability average data rate} concerns the sequence of $\{R_r(z_m)\}$ instead of $\{R_r(t_k)\}$. This expresses that the average value of bits of all the \emph{successful} transmissions, namely $\bar R_{r,m}$ for $m=1, 2, \cdots$,  should satisfy (\ref{average stability}), instead of the average value of bits of all the transmissions attempts \emph{i.e.} $\tilde{R}_{r,k}$. In fact, even if $\tilde{R}_{r,k} >  \frac{1}{1-\frac{1}{T}-\frac{\Delta}{\tau_D}}c_r \Delta \log_2e$, it is still possible that $\bar{R}_{r,m} \le  \frac{1}{1-\frac{1}{T}-\frac{\Delta}{\tau_D}}c_r \Delta \log_2e$ and instability may occur. 

In practice, the proposition above can be satisfied by computing the number of bits online so that stability can be guaranteed. For example, the coding systems can pre-compute the number of bits right before each transmission attempt such that if the transmission attempt succeeds then Proposition \ref{stability average data rate} holds. However, this implementation may lead to a larger average bit rate associated with the transmission attempts. Consider the scenario when the communication devices attempt to transmit a large number of bits to make Proposition \ref{stability average data rate} hold, but the DoS is present. Then the communication devices would attempt to send a large number of bits with each packet subsequently. Note that in this case, attempts of constantly transmitting a large number of bits are needed since it is not possible to predict the next $z_m$. If one fails the coming $z_m$, Proposition \ref{stability average data rate} maybe violated and instability may occur.

\section{Time-varying bit rate}
In this section, we aim at designing a time-varying bit-rate protocol, which preserves a comparable level of resilience against DoS while promoting the possibility of saving bits when the attack levels are low. Recalling $\bar A_r$ in Lemma \ref{Second transformation}, we equip the quantization systems corresponding to $\bar A_r$ ($r=1, 2, \cdots, p$) with their own ``clocks". Due to the utilization of the acknowledgment-based protocol, the acknowledgments could enable the encoders to update the time sequences soon after the updates of the time sequences in the decoders. This facilitates the quantization systems to use a time-varying bit-rate protocol, in which the number of bits is predetermined before each transmission and depends on the generated time sequences. In Theorem \ref{Jordan stability condition via constant rate}, we obtained a ``standard" time-invariant bit rate, namely $R_r$. Based on the obtained $R_r$, we propose a time-varying bit-rate protocol. 

We briefly introduce the intuition of the time-varying bit-rate protocol. There are two scenarios.

\emph{Scenario 1}: If the duration of a DoS attack is short, then after the attack, a transmission with fewer bits is enough to guarantee the decay of the quantization range and there is no need to transmit a large number of bits. Actually, it is this mechanism that saves bits compared with the time-invariant bit-rate protocol. 

\emph{Scenario 2}:
If the duration of a DoS attack is long, then the quantization systems may not be able to obtain the decay of the quantization range with one transmission, even by applying the maximum bit rate. 
Confronted by this problem, the transmitter attempts to send packets with the maximum number of bits, \emph{e.g.} $R_r$, for longer time until the quantization range is restored to the level smaller than the one before the ``long-time DoS" occurs. 

In the time-varying bit-rate protocol, both the encoding and decoding systems consist of two major parts: \emph{A) bit-computing parts} and \emph{B) coding parts}. The bit-computing parts pre-determine the number of bits for encoding and decoding (before each transmission attempt instant $t_k$), and the coding parts are responsible for encoding or decoding signals at $t_k$, by applying the pre-determined number of bits.  

\subsection{Bit-computing parts}

The bit-computing parts mainly generate sequences of time instants and then based on the time sequences, they pre-determine the number of bits for the transmission attempts. Note that both the encoding and decoding systems are equipped with the identical bit-computing parts and coding parts. 

We first introduce sequences of time instants generated by the bit-computing parts in the encoding and decoding systems, \emph{i.e.} $\{s_g ^ r\}=\{s_0 ^r, s_1 ^r, s_2^r,\cdots\} \subseteq \{z_m\} $ with $r=1, 2, \cdots, p$ and $g\in \mathbb{Z}_0$. In Section \uppercase\expandafter{\romannumeral 4}. C, we will show that the state corresponding to $\bar A_r$ strictly decays along $\{s_g ^ r\}=\{s_0 ^r, s_1 ^r, s_2^r,\cdots\}$ with $r=1, 2, \cdots, p$. In particular, we have
\begin{align}\label{s_k}
\left\{
\begin{array}{l}
s_g   ^r   = \min \{ z_m > s_{g-1}^ r\, | \frac{    e^{c_r(z_m-s_{g-1}^r)}    }{(2^{R_r})^{    T_S (s_{g-1} ^ r , z_m)}    }  <1      \}     \\
s_0 ^ r = z_0
\end{array}
 \right.
\end{align}
where $R_r$ satisfies Theorem \ref{Jordan stability condition via constant rate}. Note that due to the acknowledgments, $\{s_g^r\}$ can be synchronized in the encoding and decoding systems. Here abusing the notation, $ T_S (s_{g-1} ^ r, z_m) $ represents the number of successful transmissions during $]s_{g-1}^r, z_m]$. Since the number of successful transmissions during $]s_{g-1}^r, z_m]$ and $[s_{g-1}^r, z_m[$ are the same, then Lemma \ref{Lemma T} is still valid when we refer to $]s_{g-1}^r, z_m]$. By applying Lemma \ref{Lemma T} we have
\begin{align}\label{existence of s_k}
\frac{       e^{c_r(z_m-s_{g-1}^r)}    }{(2^{R_r})^{  T_S (s_{g-1} ^ r , z_m)}    }  \le \theta_r  (\alpha_r )^{z_m - s_{g-1}^r}.   
\end{align}
Note that $\alpha_r <1$ (in (\ref{alpha i})) and $\theta_r$ is finite if Theorem \ref{Jordan stability condition via constant rate} holds.
Then there always exists the smallest and finite $z_m> s_{g-1} ^r$ such that 
$\frac{ e^{c_r(z_m-s_{g-1}^r)}    }{(2^{R_r})^{  T_S (s_{g-1} ^ r , z_m)}    }  \le \theta_r  (\alpha_r )^{z_m - s_{g-1}^r}  < 1 $. Hence according to (\ref{s_k}), we have that $s_g ^r -s_{g-1}^r$ is finite when Theorem \ref{Jordan stability condition via constant rate} holds.

The pre-determined number of bits in the encoding and decoding systems for the transmission attempts follows
	\begin{align}\label{time varying bits allocation}
	&\mathcal R_l(t_k^-) \nonumber \\ 
	=\,\,& R_r(t_k^-) \nonumber\\
	=\,\,&
	\left \{
	\begin{array}{ll}
	\min \{ \underline {R_r}(t_k^-), R_r\} , & \text{if}\,  \frac{       e^{c_r(t_k^- -s_{g-1}^r)}    }{(2^{R_r})^{  T_S (s_{g-1} ^ r , t_k^ -)+1 }   }  <1  \\
	R_r, & \text{otherwise}
	\end{array}
	\right. 
	\end{align} 
where $\underline {R_r}(t_k^-): = \lceil w_r (t_k^- - s_{g-1}^ r)\log_2 e \rceil $, $t_k\in ]s_{g-1}^r, s_g ^r]$, $  w_r \in \mathbb{R}_{> c_r}$, $R_r$ satisfies Theorem \ref{Jordan stability condition via constant rate} and $r$ is the index of $\bar A_r$ that the $l$-th element corresponds to.

Note that $\mathcal R_l(t_k) = R_r(t_k) = R_r(t_k ^-)$ and if $t_k$ is a successful transmission instant such that $t_k = z_m$, then $T_S (s_{g-1} ^ r , t_k^-)+1 = T_S (s_{g-1} ^ r , z_m)$. By $T_S (s_{g-1} ^ r , t_k^-)+1$, it simply means that before the real transmission attempt at $t_k$, the bit-computing parts first estimate $\frac{ e^{c_r(t_k^ - -s_{g-1}^r)}    }{(2^{R_r})^{  T_S (s_{g-1} ^ r , t_k^-)+1} }$ by assuming that $t_k$ would be a successful transmission instant. If it estimates that by using $R_r$ bits at $t_k$, the system would have $\frac{ e^{c_r(t_k^- -s_{g-1}^r)}    }{(2^{R_r})^{  T_S (s_{g-1} ^ r , t_k^- )+1} } < 1$, then according to (\ref{time varying bits allocation}) by using $\min\{\underline {R_r}(t_k), R_r\}$ bits at $t_k$, the system would still have
\begin{align}
&\frac{ e^{c_r(t_k^- -s_{g-1}^r)}    } {2^{R_r  T_S (s_{g-1} ^ r , t_k^- ) + \min\{\underline {R_r}(t_k), R_r\}  }}   \nonumber\\
\le \,\, & \frac{ e^{c_r(t_k^- -s_{g-1}^r)}    } {2^{R_r  T_S (s_{g-1} ^ r , t_k^- )} \min\{e^{w_r(t_k^--s_{g-1}^r)}, 2^{R_r}\}} \nonumber\\
<\,\,& 1
\end{align}
which in turn implies the decay of the quantization range (see Section \uppercase\expandafter{\romannumeral 4}. C).
Since $	\min \{ \underline {R_r}(t_k^-), R_r\}  \le R_r$, we achieve the possibility of the reduction of bits.    
For the ease of visualization, the evolution of $\frac{ e^{c_r(z_m-s_{g-1}^r)}    }{(2^{R_r})^{  T_S (s_{g-1} ^ r , z_m)}    } $ ($c_r > 0$), $\{s_g ^r\}$ and the applied number of bits are exemplified in Figure \ref{time varying picture}.

\begin{figure}[t]
	\begin{center}
		\includegraphics[width=0.5 \textwidth]{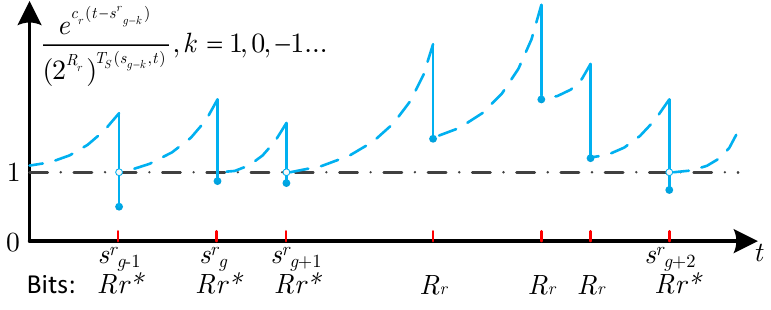} \\
		\linespread{1}\caption{ The evolution of $\frac{ e^{c_r(z_m-s_{g-1}^r)}    }{(2^{R_r})^{  T_S (s_{g-1}^r  , z_m)}} $ and $\{s_g ^r\}$. The instants of successful transmissions ($z_m$) are indicated by the red lines, at which there are the values of $\frac{ e^{c_r(z_m-s^r_{g-1})}    }{(2^{R_r})^{  T_S (s_{g-1}^r  , z_m)}} $ being indicated by the blue solid dots and the number of applied bits ($R_r$ or $R_r^*:=\min \{R_r, \underline {R_r}\}$).  Among the instants of successful transmissions, we highlight the sequence $\{s_g^r\}$, at which the values of $\frac{ e^{c_r(z_m-s_{g-1}^r)}    }{(2^{R_r})^{  T_S (s_{g-1}^r  , z_m)}    } $ are smaller than 1. The dashed blue curves represent the evolution due to $e^{c_r t}$ and the solid blue lines represent their drops due to the successful transmissions.  
		} \label{time varying picture}
	\end{center}
\end{figure}

\subsection{Coding parts}

In the last part, we obtain the pre-determined number of bits applied to each transmission attempt. By applying such a number of bits, the coding parts consisting of the quantizers (\ref{time varying quantizer})-(\ref{time varying quantizer 2}), the impulsive system of quantization range (\ref{time varying J}) and the predictor (\ref{time varying predictor}) are introduced in this part. 

The uniform quantizer in (\ref{uniform quantizer}) with time-varying bits is
\begin{align} \label{time varying quantizer}
q_{\mathcal R_l(t_k)} (\chi_l) : = 
\left \{ \begin{array}{ll} 
\frac{\lfloor 2^{\mathcal R_l(t_k)-1} \chi_l \rfloor + 0.5}{ 2^{\mathcal R_l(t_k)-1} }, &  \quad  \textrm{if } -1 \le \chi_l < 1  \\ 
1- \frac{0.5}{2^{\mathcal R_l(t_k)-1}}, & \quad  \textrm{if } \chi_l=1
\end{array} \right. 
\end{align} 
if $\mathcal R_l (t_k)> 0$,
and in particular,
\begin{align}\label{time varying quantizer 2}
q _{\mathcal R_l(t_k)} (\chi_l) =0
\end{align}
if $\mathcal R_l(t_k) = 0 $. Likewise, the property
\begin{align}
\left| e_l - j_l q_{\mathcal R_l(t)}\left(\frac{e_l}{j_l}\right) \right|\le \frac{j_l}{2^{\mathcal R_l(t)}}, \,\, \text{if}\,\, \frac{|e_l|}{j_l} \le 1
\end{align}
still holds in the time-varying-bit scenario.

Note that in the time-varying bit-rate design, the coding parts are the same as in (\ref{Jordan predictor})-(\ref{Jordan initial condition}) except that $\Phi(t)$ in (\ref{Jordan case Phi}) and $H$ in (\ref{Jordan case H}) should be changed into the time-varying-bit forms $\bar \Phi(t)$ and $H(t)$, respectively. In $\bar \Phi(t)$, the number of bits are time-varying instead of time-invariant, and we have
\begin{align}\label{time varying Phi}
\bar \Phi(t)=\left [  \begin{array}{ccc}
\bar \phi_1(t) \\
\vdots\\
\bar \phi_{n_x}(t)
\end{array}  \right]=\left [  \begin{array}{ccc}
j_1 (t) q_{\mathcal  R_1(t)}( \chi_1(t) ) \\
\vdots\\
j_{n_x} (t) q_{\mathcal R_{n_x}(t)}( \chi _{n_x}(t) )
\end{array}  \right]
\end{align}
Meanwhile, $H(t)$ is given by
\begin{align}\label{time varying H}
H(t) = \text{diag}(2^{-R_1(t)}I_1, 2^{-R_2(t)}I_2, \cdots, 2^{-R_p(t)}I_p)
\end{align}
and we let $H_r(t):= 2^{-R_r(t)}I_r $.
Then we have the impulsive system for obtaining the quantization range
\begin{align}\label{time varying J}
\left\{
\begin{array}{ll}
\dot J(t) = \bar A J(t), & \text{if} \,\, t\ne z_m \\
J(t)= H(t) J(t),  & \text{if} \,\, t= z_m
\end{array}
\right.
\end{align}
At last, the predictor in the time-varying bit-rate design is given by
\begin{align}\label{time varying predictor}
\left\{ \begin{array}{ll} 
\dot{\hat x}(t) = \bar{A} \hat x(t) + \bar B(t) u(t), & t\ne z_m \\
\hat x(t) = \hat x(t^-) - \bar \Phi(t^-), & t=z_m
\end{array}\right.
\end{align}
where $u(t)= \bar{K} \hat x(t)$ and (\ref{Jordan initial condition}) holds. By applying a very similar analysis as in Section \uppercase\expandafter{\romannumeral 3}. C, one could see that there is no over-flow problem of the quantization systems under the time-varying bit-rate protocol.

In view of (\ref{s_k}), (\ref{time varying bits allocation}) and the coding parts, the mechanism of the time-varying bit-rate protocol can be outlined as
\begin{itemize}
	\item[1)] Set $s_{0}^r = z_0$.
	\item[2)] Let $t_k$ be the transmission attempt instant after $ s_{g-1}^r$ such that $  s_{g-1}^r < t_k \le s_{g}^r$. At $t_k^-$, the bit-computing parts in the encoding and decoding systems calculate $\frac{ e^{c_r(t_k^--s_{g-1}^r)}    }{(2^{R_r})^{  T_S (s_{g-1} ^ r , t_k^-)+1 } }$.
	\begin{itemize}
		\item[2.1)] If $\frac{ e^{c_r(t_k^- -s_{g-1}^r)}    }{(2^{R_r})^{  T_S (s_{g-1} ^ r , t_k^- )+1} } < 1$, then in view of (\ref{time varying bits allocation}), we have $\mathcal R_l(t_k^-) = R_r(t_k ^-) = \min \{ \underline R_r (t_k ^-) , \, R_r \}$, and $R_r(t_k) = R_r(t_k^-)$ in the bit-computing parts.
		\begin{itemize}
			\item[2.1.1)] If the transmission succeeds ($t_k = z_m$), update $J$ in (\ref{time varying J}), $\hat x$ in (\ref{time varying predictor}) by using the number of bits determined in 2.1), and update $s^r _{g} = z_m$ in light of (\ref{s_k}) in both the encoding and decoding systems. Then we are at 2) with $s^r _{g-1}$ becoming $s^r _{g}$ and wait for the next $t_k$ and repeat 2).
			\item[2.1.2)] If the transmission fails, wait for the next $t_k$ and repeat 2).
		\end{itemize}
		\item[2.2)] If $\frac{ e^{c_r(t_k^ - -s_{g-1}^r)}    }{(2^{R_r})^{  T_S (s_{g-1} ^ r , t_k^-)+1} } \ge 1 $, then according to (\ref{time varying bits allocation}), we have $\mathcal R_l(t_k^-) = R_r(t_k ^-) = R_r $, and $R_r(t_k) = R_r(t_k^-)$.
		\begin{itemize}
			\item[2.2.1)] If the transmission succeeds, update $J$ in (\ref{time varying J}), $\hat x$ in (\ref{time varying predictor}) by using the number of bits determined in 2.2) in both the encoding and decoding systems. Then wait for the next $t_k$ and repeat 2).
			\item[2.2.2)] If the transmission fails, wait for the next $t_k$ and repeat 2).
		\end{itemize}
	\end{itemize}
\end{itemize}

\subsection{Stability analysis}
For the ease of conveying the ideas, we focus our analysis on the dynamics corresponding to $\bar A_r$.
From (\ref{time varying J}), it is easy to obtain that 
\begin{align}
J_r (s_g ^ r) = \bar P_r(s_{g-1}^r, s_g ^r) J_r(s_{g-1}^r)
\end{align}
with 
\begin{align}
\bar P_r (s_{g-1}^r, s_g ^r) = e^{\bar A_r (s_{g}^r - s_{g-1}^r)} \prod_{s_{g-1}^r <z_m \le s_g ^ r } H_r(z_m) 
\end{align}
where $H_r(z_m) = 2^{-R_r(z_m)} I_r$ and $J_r$ is the subset of $J$ corresponding to $\bar A_r$. Since $e^{\bar A_r (s_{g}^r - s_{g-1}^r)}= U_r(s_{g}^r - s_{g-1}^r) $ is upper-triangular whose eigenvalues equal to $e^{c_r(s_g ^ r- s_{g-1} ^r)}$, and $\prod_{s_{g-1}^r < z_m \le s_g ^r } H_r(z_m)$ is diagonal, it is easy to obtain the eigenvalues of $\bar P_r(s_{g-1}^r, s_g ^r)$ such that 
\begin{align}\label{time varying eigenvalue}
\lambda_r(\bar P_r(s_{g-1}^r, s_g ^r)) = \frac{e^{c_r(s_g ^ r - s_{g-1 }^r)}}{\prod_{s_{g-1}^r < z_m \le s_g ^r} 2^{R_r(z_m)}}
\end{align}
Iteratively, it is easy to verify that 
\begin{align}\label{time varying J_r}
J_r(s_g ^r) = \prod_{k=1}^g   \bar P_r(s_{k-1}^r, s_k ^r) J_r(s_0 ^r) 
\end{align}

Recall the definition of $\{s_g ^r\}$ in (\ref{s_k}). The next lemma concerns the convergence of $J_r(s_g ^ r)$.

\begin{lemma}\label{lemma J_r}
	Consider the impulsive system as in (\ref{time varying H})--(\ref{time varying J}) and the DoS attacks in Assumptions 1 and 2 satisfying $\frac{1}{T}+ \frac{\Delta}{\tau_D}<1$ with $\Delta$ being the sampling interval of the network as in (\ref{periodic transmission interval}).  The time-varying-bit quantizer is given by (\ref{time varying quantizer})-(\ref{time varying quantizer 2}), where $R_r$ satisfies Theorem \ref{Jordan stability condition via constant rate} and $w_r > c_r$. Then, all the elements in the column vector $J_r(s_{g}^r)$ converge to zero as $g \to \infty$ with $r=1, 2, \cdots, p$.
\end{lemma}

\emph{Proof.} In the proof, we would like to show that the eigenvalues of $\bar P_r(s_{g-1}^r, s_g ^r)$ satisfy $\lambda_r(\bar P_r(s_{g-1}^r, s_g ^r))<1$ for $g\in \mathbb{Z}_1$.

Let $z_m^*$ denote the first successful transmission instant after $s_{g-1}^r$. If $\frac{e^{c_r(z_m^ {*-} - s_{g-1} ^ r)}}{(2^{R_r})^{T_S(s_{g-1}, z_m ^{*-} )+1} } = \frac{e^{c_r(z_m^ * - s_{g-1} ^ r)}}{(2^{R_r})^{T_S(s_{g-1}, z_m ^*)} } =  \frac{e^{c_r(z_m^ * - s_{g-1} ^ r)}}{2^{R_r}} < 1$, then according to (\ref{time varying bits allocation}), $\min \{\underline R_r(z_m^*), R_r\}$ would be applied for coding. 
Hence based on (\ref{time varying eigenvalue}), we obtain that 
\begin{align}\label{time varying short eigenlave}
\lambda_r(\bar P_r(s_{g-1}^r, z_m ^*)) &= \frac{e^{c_r(z_m ^* - s_{g-1 }^ r)}}{\prod_{s_{g-1}^r < z_m \le z_m ^*} 2^{R_r(z_m)}}
\nonumber\\
&= \frac{e^{c_r(z_m ^* - s_{g-1} ^ r)}}{ 2^{R_r(s_g ^ r)}} \nonumber\\
&= \frac{e^{c_r(z_m ^* - s_{g-1} ^ r)}}{ 2^{\min\{\underline{R_r}(z_m ^*), R_r\}                 }} \nonumber\\
&<1
\end{align}
where the inequality is implied by the hypothesis. Meanwhile, we see that such $z_m ^*$ qualifies (\ref{s_k}) and hence $s_g ^r 
= z_m ^*$. One obtains that $\lambda_r(\bar P_r(s_{g-1}^r, s_g ^ r))< 1$.  

If $\frac{e^{c_r(z_m^ {*-} - s_{g-1} ^ r)}}{(2^{R_r})^{T_S(s_{g-1}, z_m ^{*-} )+1} } = \frac{e^{c_r(z_m^ * - s_{g-1} ^ r)}}{(2^{R_r})^{T_S(s_{g-1}, z_m ^*)} } =  \frac{e^{c_r(z_m^ * - s_{g-1} ^ r)}}{2^{R_r}} \ge 1$, then according to (\ref{time varying bits allocation}), the coding systems would apply $R_r$ bits at $z_m ^*$ and during $]z_m ^*, z_m^{**}]$ where $z_m ^{**} > z_m ^{*}$. Therefore, according to (\ref{time varying eigenvalue}), we obtain that 
\begin{align}\label{timn varying long eigenvalue}
\lambda_r(\bar P_r(s_{g-1}^r, z_m ^{**})) 
&= \frac{e^{c_r(z_m^{**}  - s_{g-1 }^ r)}}{\prod_{s_{g-1}^r < z_m \le z_ m  ^{**} } 2^{R_r(z_m)}}
\nonumber\\
&= \frac{e^{c_r(z_m ^ * - s_{g-1 }^ r)}}{2^{R_r}} \frac{e^{c_r(z_m^{**}   - z_{m }^ *)}}{(2^{R_r})^{T_S(z_m^*,  z_{m}^{**} )}}  \nonumber\\
									   &=  \frac{e^{c_r(z_m^{**}  - s_{g-1 }^ r)}}{(2^{R_r})^{T_S(s_{g-1}^r,  z_{m}^{**})}} 
\nonumber\\
&<1
\end{align}
where the rationale of the inequality and the existence of such $z_m^{**}$ have been discussed in (\ref{existence of s_k}) and the discussion thereafter. Hence such $z_m^{**}$ is denoted by $s_g ^r$ and we have $\lambda_r(\bar P_r(s_{g-1}^r, s_g ^r)) < 1$.

In either case, we have shown that $\lambda_r(\bar P_r(s_{g-1}^r, s_g ^r)) <1$, which implies that $\{\bar P_r(s_{g-1}^r, s_g ^r)\}_{g\in \mathbb{Z}_1}$ is a sequence of stable matrices and there exist finite $C^r _4$ and $0<\beta_r<1$ such that $\|J_r(s_g ^ r)\| \le C_4 ^ r (\beta_r) ^{g} \|J_r(s_0^r)\|$ in view of (\ref{time varying J_r}). Therefore, one can infer that $J_r(s_g ^r) \to 0$ when $g \to \infty$ with $r=1, 2, \cdots, p$. This completes the proof.
\qedp

In view of the dynamics of $J(t)$, we have
\begin{align}
\|J_r(t)\| \le C_4 ^ r     e^{v_r(t-s_{g-1} ^{r})}         (\beta_r) ^{g-1} \|J_r(s_0^r)\|
\end{align}
where $ s_{g-1} ^{r}< t \le s_g ^r $ and $v_r=\max \{0, \bar v_r\}$ with $\bar v_r =\lambda_{\max}(\frac{\bar A_r + \bar A_r ^{\text{T}}}{2})$.
Since $0<\beta_r<1$ and $s_{g}^r - s_{g-1}^r$ is finite, one knows that there exist finite $\bar \gamma_r$ and $\bar \mu_r<0$ such that $\|J_r(t)\|   \le \bar \gamma_r e^{\bar \mu_r(t-s_0^r)} \|J_r(s_0 ^r)\|$. This implies that there exists finite $\bar \gamma $ and $\bar \mu <0$ such that $\|J(t)\| \le \bar \gamma e^{\bar \mu (t-z_0)}\|J(z_0)\|$ with $r=1, 2, \cdots, p$ by noticing that $s_0 ^r = z_0$. Since $|e|$ is upper bounded by $J$, we obtain that $\|e(t)\| \le \|J(t)\| \le \bar \gamma e^{\bar \mu (t-z_0)} \|J(z_0)\|$.

\begin{theorem}\label{time-varying theorem}
		Consider the process (\ref{vector system}) with control action (\ref{time varying Phi})-(\ref{time varying predictor}) under the transmission policy in (\ref{periodic transmission interval}). Suppose the DoS attacks characterized as in Assumptions 1 and 2 and satisfy $\frac{1}{T}+ \frac{\Delta}{\tau_D}<1$. The transmitted signals are quantized by the time-varying-bit quantizer (\ref{time varying quantizer})-(\ref{time varying quantizer 2}) where $R_r$ satisfies Theorem \ref{Jordan stability condition via constant rate} and $w_r > c_r$. Then the closed-loop system is exponentially stable.
\end{theorem}

\emph{Proof.} Since $e(t)$ exponentially converges to the origin in view of Lemma \ref{lemma J_r} and the discussion thereafter, following the very similar calculation as in the proof of Theorem \ref{Jordan stability condition via constant rate}, one can obtain the exponential stability of the closed-loop system.
\qedp

\begin{remark}\label{slow convergence speed}
It is worth mentioning that the reduction of bits is achieved by sacrificing the decay rate of the system, \emph{i.e.} the system converges in a slower rate compared with the one under the time-invariant bit-rate protocol. This is due to the fact that in the absence of DoS attacks or after short-duration DoS attacks, the time-invariant bit-rate protocol is able to apply $R_r$ bits, while the time-varying bit-rate protocol can only apply $\min \{\underline {R_r} (t_k) ,   R_r \}$ bits (\emph{cf.} Figure \ref{time varying picture}).   \qedp
\end{remark}

\begin{remark}
One sees that the system under control is stable if one chooses $w_r$ and $R_r$ properly. It is easy to make the design parameter $w_r > c_r$ and hence we omit the influence of it. Then $R_r$ is the only parameter affecting the robustness of the system under time-varying bit-rate protocol. In view of Remark \ref{Remark 4}, we characterize the system's robustness under time-varying bit-rate protocol such that if (\ref{Jordan DoS condition for constant data rate}) holds, then the stability of the closed-loop system can be preserved.
\qedp
\end{remark}

\section{Numerical example}
For simplicity, we consider a process that is in Jordan form and taken from \cite{forni}. The system to be controlled
is open-loop unstable and is  
characterized by the matrices 
\begin{eqnarray} \label{eq:system_example}
A=\tilde{A}=\bar A =\left[ \begin{array}{cc} \, 1 \, & \, 1 \, \\ 0 \, &  \, 1 \end{array} \right], \quad 
B=\tilde{B} = \bar B(t) = \left[ \begin{array}{cc} \, 1 \, & \, 0 \, \\ 0 \, &  \, 1 \end{array} \right]
\end{eqnarray} 
The state-feedback matrix is given by
\begin{eqnarray} \label{eq:controller_example}
K =\bar K(t) = \left[ \begin{array}{cc} -2.1961 & -0.7545 \\ -0.7545 & -2.7146 \end{array} \right]
\end{eqnarray} 
The eigenvalues of $A$ are $1$.

The network transmission interval is given by $\Delta = 0.1$s. We consider a 
sustained DoS attack with variable period and duty cycle, generated randomly.
% with values $\tau_D>10$ and $T>1.25$
% which causes $\sim80\%$ of transmission failures. 
Over a simulation horizon of $20$s, the DoS 
signal yields $|\Xi(0,20)|=15.52$s and $n(0,20)=20$. 
This corresponds to values (averaged over $20$s) 
of $\tau_D\approx0.96$ and $T\approx1.29$, and $\sim80\%$ of transmission failures. 
It is simple to verify that 
\begin{eqnarray}
\frac{\Delta}{\tau_D} + \frac{1}{T} \approx 0.8793
\end{eqnarray}

According to Theorem \ref{Jordan stability condition via constant rate}, we obtain that 
\begin{align}
R_1 > \frac{c_r \Delta}{1-\frac{1}{T}-\frac{\Delta}{\tau_D}}\log_2 e = 1.1953
\end{align}
Then we select $R_1=2$. Meanwhile, considering the choice of $w_1$ satisfying $ w_1 > c_1 =1$, we let $w_1=2$ for the time-varying bit-rate protocol.
% such that
%\begin{align}
%\bar{R_i}(t_k) =  \min \{\lceil 2 (t_k-s^i_{r-1}) \log_2e  \rceil, 2\},
%\end{align}
%see (\ref{time varying quantizer})-(\ref{time varying bits allocation}) and the discussion thereafter.
The simulation results of $x(t)$ are shown in Figure \ref{simulation}. We see that $x(t)$ converges to the equilibrium in both protocols. In particular, the state in the bottom picture converges with a slightly lower speed. This is due to the fact that in the absence of DoS or after a ``short-duration" DoS attack, the network transmits fewer bits (\emph{cf.} Remark \ref{slow convergence speed}). This can be observed from Figure \ref{simulation2}. One could see that the convergence of $J(t)$ under time-varying bit-rate protocol shown in the middle picture of Figure \ref{simulation2} (the numbers of bits applied in the time-varying bit-rate protocol are shown in the bottom picture of Figure \ref{simulation2}) is slower than the one under time-invariant bit-rate protocol as shown in the top picture of Figure \ref{simulation2}.

In fact, the obtained values of bit rate are conservative in the time-invariant bit-rate protocol. The stability can be still preserved at the lower rate with $R_1=1$ under the same pattern of DoS attacks. One factor contributing to the conservativeness is that the actual number of successful transmissions is much larger than the theoretical value computed in Lemma \ref{Lemma T}. 

From another viewpoint, if the data rate of the channel is pre-selected as $R_1=2$, the closed-loop system should be stable under the attacks in this example since the DoS parameters satisfy $\frac{1}{T}  + \frac{\Delta}{\tau_D}  \approx 0.8793 < 1-\frac{c_1\Delta\log_2e}{R_1}=0.9279$.

\begin{figure}[t]
	\begin{center}
		\includegraphics[width=0.4 \textwidth]{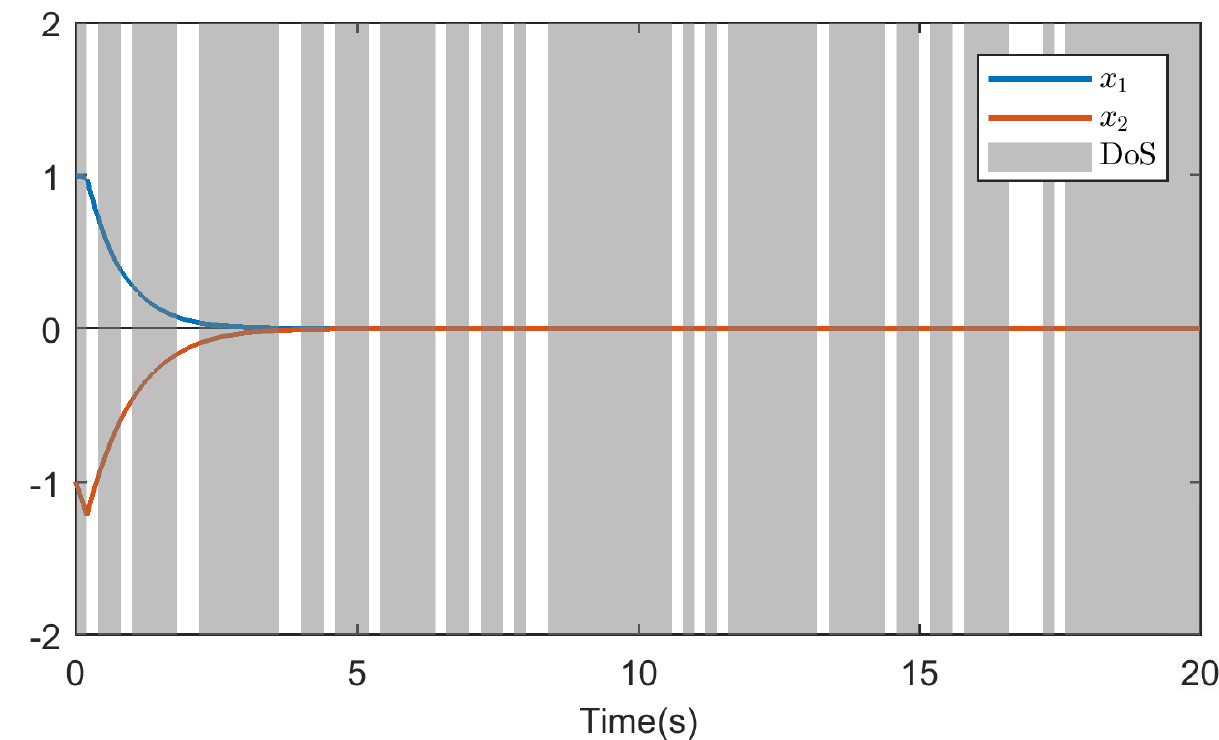} \\
		\includegraphics[width=0.4 \textwidth]{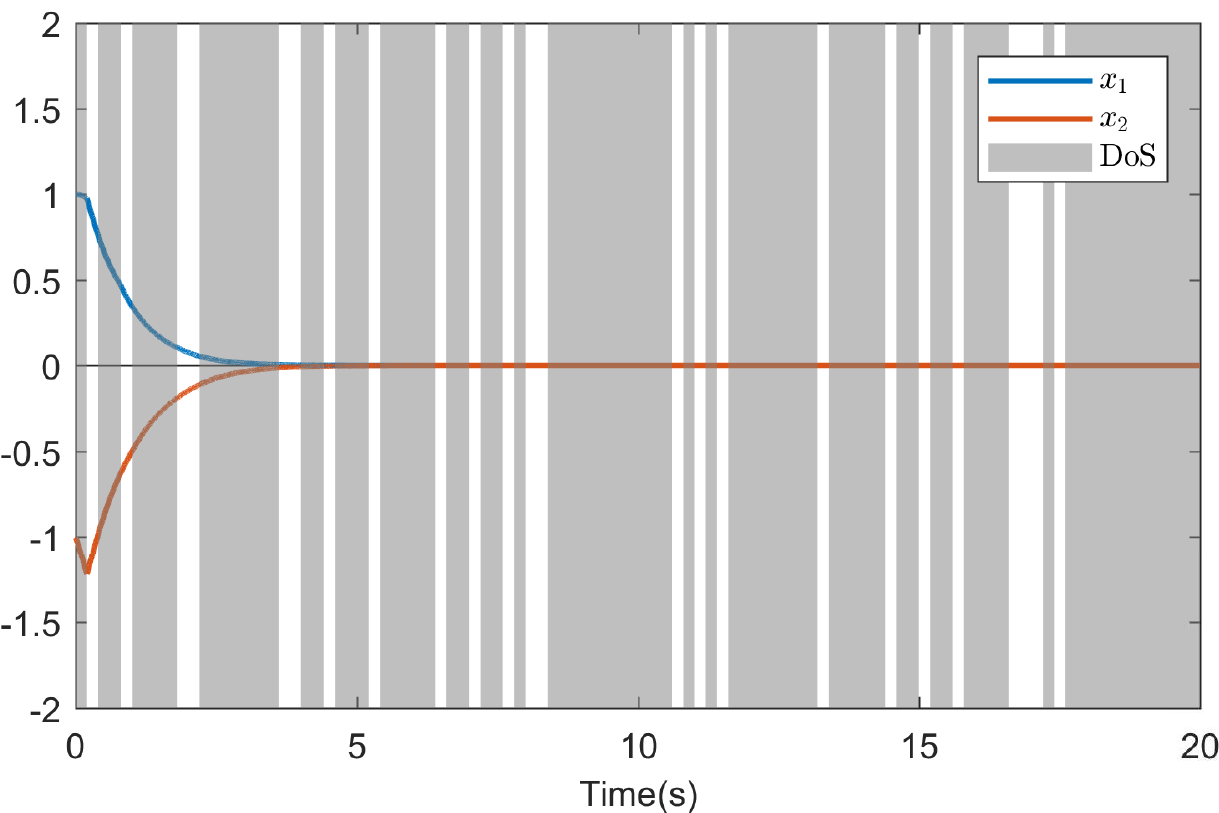} \\
		\linespread{1}\caption{Simulation plots of $x(t)$. Top picture--State under time-invariant bit-rate protocol; Bottom picture--State under time-varying bit-rate protocol.   
		} \label{simulation}
	\end{center}
\end{figure}

\begin{figure}[t]
	\begin{center}
		\includegraphics[width=0.4 \textwidth]{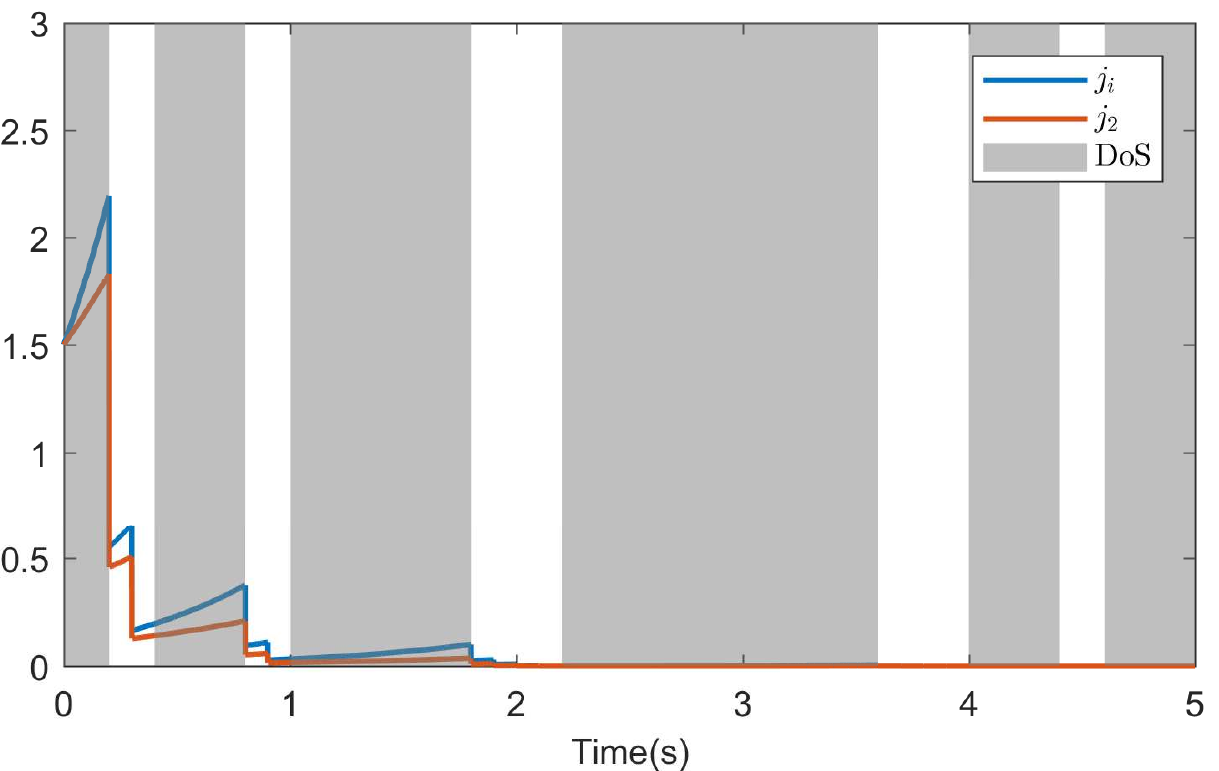} \\
		\includegraphics[width=0.4 \textwidth]{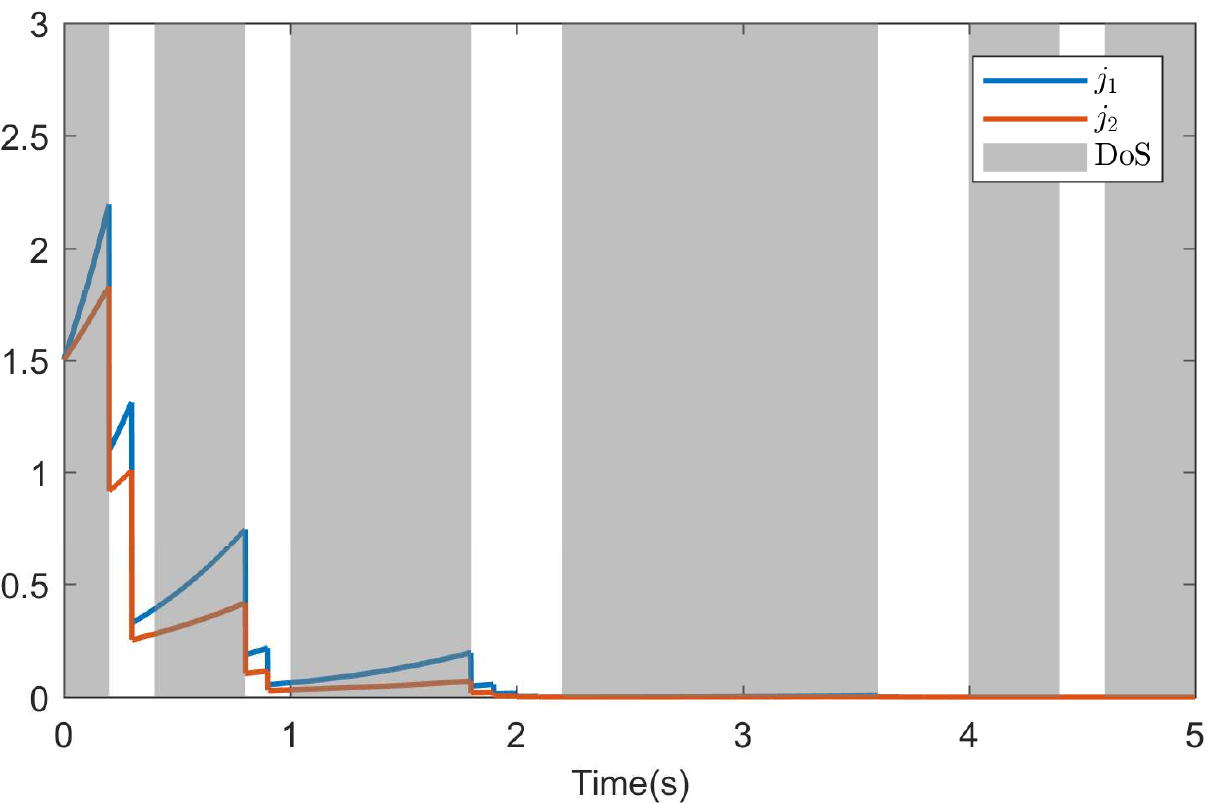} \\
		\includegraphics[width=0.4 \textwidth]{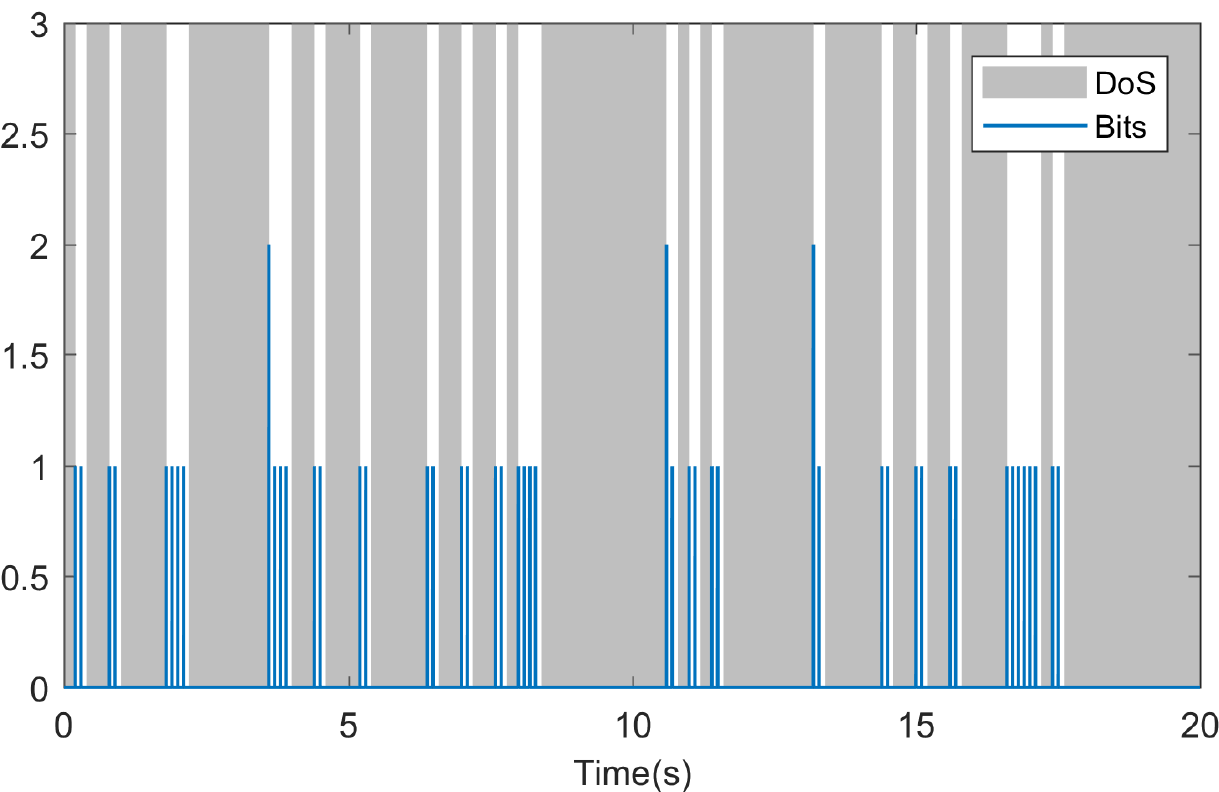} \\
		\linespread{1}\caption{Simulation plots. Top picture--$J(t)$ under time-invariant bit-rate protocol, where the total number of bits associated with all transmission attempts over 20s is 800; Middle picture--$J(t)$ under time-varying bit-rate protocol, where the total number of bits associated with all transmission attempts over 20s is 452; Bottom picture--successful transmitted bits under time-varying bit-rate protocol.
		} \label{simulation2}
	\end{center}
\end{figure}

\section{Conclusions}

We investigated the data rate problem for stabilizing control of a networked control system under limited bandwidth and Denial-of-Service attacks. It was shown that the sufficient condition of bit rate for stabilization depends on the unstable eigenvalues of the dynamic matrix of the process as well as the DoS parameters. Furthermore, the design of time-varying bit-rate protocol is proven to be effective in saving bits meanwhile maintaining the comparable resilience as the one under time-invariant bit-rate protocol. 
It is emphasized that the results of the paper clearly indicate the trade-offs between the amount of transmitted data and the robustness against DoS attacks. In particular, the approach is in accordance with the recent studies on the minimum data rate control problems.

In the future, disturbance and noise might be taken into consideration. Moreover, the analysis in this paper can be possibly applied to achieve the corresponding bit-rate bounds under the different packet-drop models considered in \cite{8057735}.

\appendix
\emph{Proof of Lemma \ref{Second transformation}}.
Recall $\tilde{A}$ in (\ref{Jordan form}), $A_r$ in (\ref{Jordan block}) and (\ref{Jordan complex block}) representing the Jordan block associated with real and complex eigenvalues, respectively. Let 
\begin{align}\setlength{\arraycolsep}{3pt}
E(t) = \left[\begin{array}{cccc}
E_1(t) & & & \\ & E_2(t) &  & \\ & & \ddots & \\ & & & E_p(t) 
\end{array}\right]\in  \mathbb{R}^{n_x \times n_x}
\end{align}
with $p\in \mathbb{Z}_1$, where
\begin{align}\setlength{\arraycolsep}{4pt}\label{P_i real}
E_r(t) = \left[\begin{array}{cccc}
1 & & & \\ & 1 &  & \\ & & \ddots & \\ & & & 1 
\end{array}\right]\in  \mathbb{R}^{n_r \times n_r}
\end{align}
corresponds to the real eigenvalue $\lambda_r =c_r$, and  
\begin{align}\setlength{\arraycolsep}{3pt}\label{P_i}
E_r(t) = \left[\begin{array}{cccc}
\varpi _r (t) & & & \\ & \varpi _r (t) &  & \\ & & \ddots & \\ & & & \varpi _r (t) 
\end{array}\right]\in  \mathbb{R}^{2n_r \times 2n_r}
\end{align}
with 
\begin{align}
\varpi _r (t) =
\left[\begin{array}{cc}
\cos(d_r t) & \sin(d_r t) \\ -\sin(d_r t) & \cos(d_r t)  
\end{array}\right]
\end{align}
corresponds to the complex eigenvalues $\lambda_r = c_r \pm d_r i$ ($d_r \ne 0$).

Since $\bar x(t) = E(t) \tilde{x}(t)$, it is easy to verify that 
\begin{align}
\dot {\bar{x}} (t) &= E(t) \dot{\tilde{x}}(t) +\dot  E(t) \tilde{x}(t) \nonumber\\
                     &= E(t) (\tilde{A} \tilde{x}(t) + \tilde{B} u(t)) + \dot E(t) \tilde{x}(t) \nonumber\\
                     &= E(t) \tilde{A} E(t)^{-1} \bar x(t) + \dot E(t) E(t)^{-1} \bar x(t) + E(t) \tilde{B} u(t) \nonumber\\
                     &= (E(t) \tilde{A} E(t)^{-1} + \dot E(t) E(t)^{-1}) \bar x(t) + E(t) \tilde{B} u(t) .
\end{align}
Let $\bar A := E(t) \tilde{A} E(t)^{-1} + \dot E(t) E(t)^{-1}= \text{diag}(\bar A_1, \bar A_2, \cdots, \bar A_p)$ and $\bar B(t): = E(t) \tilde{B} $, where 
\begin{align}
\bar A_r := E_r(t) \tilde{A_r} E_r(t) ^{-1} + \dot E_r(t) E_r(t) ^{-1}.
\end{align}

If the eigenvalues associated with $A_r$ are real, then $E_r(t)$ is an identity matrix in (\ref{P_i real}) with order $n_r$ and hence the derivative of $E_r(t)$ is a matrix with only zero entries, which implies that
\begin{align}\label{real block}
\bar A_r &= E_r(t) \tilde{A}_r E_r(t) ^{-1}+\dot E_r(t) E_r(t) ^{-1} \nonumber\\
 &=\tilde{A_r}= \left[\begin{array}{cccc}
c_r &  1& & \\ & c_r & 1 & \\ & & \ddots & 1\\ & & &  c_r
\end{array}\right].
\end{align}

If the eigenvalues associated with $A_r$ are complex, \emph{i.e.} $\lambda_r= c_r \pm d_r i$ with $d_r \ne 0$, then $E_r(t)$ is a time-varying matrix as in (\ref{P_i}), whose derivative is not zero any longer. It is simple to verify that 
\begin{align}
E_r(t) \tilde{A_r} E_r(t) ^{-1} 
 =
 \left[\begin{array}{cccc}
D_r & I & &\\ & D_r & I &  \\ &&\ddots& I\\ &&& D_r
\end{array}\right]
\end{align}
with $E_r(t)$ being in (\ref{P_i}) and recalling that
\begin{align}
D_r = 
\left[\begin{array}{cc}
c_r & -d_r \\ d_r & c_r
\end{array}\right], \,\,
I = 
\setlength{\arraycolsep}{4pt}
\left[\begin{array}{cc} 
1 & 0 \\ 0 & 1
\end{array}\right]
\end{align}
On the other hand, we have
\begin{align}
\dot E_r(t) E_r(t) ^{-1}
=
\left[\begin{array}{cccc}
F_r &  & &\\ & F_r &  &  \\ &&\ddots& \\ &&& F_r
\end{array}\right], \,\,  \text{where}\,
F_r = 
\left[\begin{array}{cc}
0 & d_r \\ -d_r & 0
\end{array}\right].
\end{align}
Thus, 
\begin{align}\label{imaginary block}
\bar A_r 
= &E_r(t) \tilde{A} E_r(t) ^{-1} + \dot E_r(t) E_r(t) ^{-1}  \nonumber\\
= &
\left[\begin{array}{cccc}
c_r I  & I & & \\ & c_r I  & I & \\ & & \ddots & I \\ & & & c_r I 
\end{array}\right].
\end{align}

Considering the two scenarios in (\ref{real block}) and (\ref{imaginary block}), we obtain the result as in Lemma \ref{Second transformation}. This completes the proof. \qedp

\bibliographystyle{IEEEtran}
\bibliography{quantization}

\end{document}